\documentclass[twocolumn,trackchanges]{aastex631}

\usepackage{color}
\usepackage{threeparttable}
\usepackage{longtable}

\shorttitle{TDEs with Rubin}
\shortauthors{Bu{\v c}ar Bricman et al.}

\graphicspath{{./}{figures/}}

\begin{document}

\title{Rubin Observatory's Survey Strategy Performance for Tidal Disruption Events}


\correspondingauthor{Andreja Gomboc}
\email{andreja.gomboc@ung.si}

\author[0000-0001-6494-9045]{K. Bu{\v c}ar Bricman}
\affiliation{Center for Astrophysics and Cosmology, University of Nova Gorica, SI-5000 Nova Gorica, Slovenia}

\author[0000-0002-3859-8074]{S. van Velzen}
\affiliation{Leiden Observatory, Leiden University, The Netherlands}

\author[0000-0002-2555-3192]{M. Nicholl}
\affiliation{Astrophysics Research Centre, School of Mathematics and Physics, Queens University Belfast, Belfast BT7 1NN, UK
}
\affiliation{Institute for Gravitational Wave Astronomy and School of Physics and Astronomy, University of Birmingham, Birmingham B15 2TT, UK}

\author[0000-0002-0908-914X]{A. Gomboc}
\affiliation{Center for Astrophysics and Cosmology, University of Nova Gorica, SI-5000 Nova Gorica, Slovenia}
\affiliation{Department of Astronomy and the DiRAC Institute, University of Washington, 3910 15th Avenue, NE, Seattle, WA 98195, USA}


\begin{abstract}

Tidal Disruption Events (TDEs) are rare transients, which are considered as promising tools in probing supermassive black holes in quiescent galaxies. The majority of $\approx 60$ known TDEs has been discovered with time-domain surveys in the last two decades. Currently, $\approx 10$ TDEs are discovered per year, and this number will increase with the Legacy Survey of Space and Time (LSST) at Rubin Observatory. This work evaluates LSST survey strategies in view of their performance in identifying TDEs.
We assume that TDEs can be identified photometrically based on their colors, particularly $u$-band, and will be scientifically useful if we can detect the light curve peak to derive physical quantities.
We define requirements for the Rubin light curves needed to achieve this (detections pre-peak, post-peak, in different bands to measure colour). We then 
inject model light curves into the Operations Simulator, and calculate the fraction of TDEs passing our requirements for several strategies. 
We find that under the baseline strategy, $\approx 1.5$\% of simulated TDEs fulfil our detection criteria, while this number increases when more time is devoted to $u$-band observations. An ideal observing strategy for photometric identification of TDEs would have longer $u$-band exposures, which should not come at the expense of fewer $u$-band visits. A filter distribution weighted towards more observing time in bluer bands, intra-night visits in different filters, and strategies with frequent sampling leading to higher quality light curves are preferred. We find that these strategies benefiting TDE science do not impact significantly other science cases.

\end{abstract}


\section{Introduction} 
\label{sec:intro}

Tidal Disruption Events (TDEs) are relatively rare transients ($\sim 100$\,Gpc$^{-3}$\,yr$^{-1}$; \citealt{vanVelzen:2018ApJ, Stone2020SSRv}) which occur when a star is tidally disrupted by a supermassive black hole (SMBH), causing a bright flare of light from the nucleus of a galaxy which may be otherwise quiescent \citep{Rees:1988bf,Evans:1989qe}. Their observed emission depends on parameters such as the black hole mass and spin, the stellar mass, radius and structure, and the pericenter distance of the encounter \citep[e.g][]{Kochanek:1993cm, Gomboc:2005wu, Guillochon:2012uc, Mockler:2018xne}. TDEs hence offer a unique opportunity to measure masses and spins of dormant SMBHs and to probe stellar populations and dynamics in galactic centers.

A few dozen TDEs have been observed to date, with the majority discovered by wide-field optical surveys in the past decade \citep[see][for recent reviews]{vanVelzen:2020SSRv, Gezari2021ARAA}. TDE light curves tend to show a rise on a timescale of a month, followed by a decay from peak often consistent with $t^{-5/3}$ as expected from bound matter returning towards the SMBH, and a late-time power-law decline consistent with accretion disk emission. However, several TDEs with faster rising and fading times that deviate from the $t^{-5/3}$ decline have also been found. The peak UV/optical absolute magnitudes cluster around $-20$ mag, but TDEs generally sample a couple of orders of magnitude spread in peak luminosity. Another characteristic typicall to TDEs is their blue color indicating high black body temperatures ($T_{{BB}} \approx 2 \times 10^4$ K). 

Many uncertainties remain in our understanding of TDEs. While X-ray emission detected in some TDEs can be explained by an accretion disk, our knowledge about the origin of the UV/optical emission is incomplete. It has been suggested that the UV/optical light originates either from a reprocessing layer, where X-ray photons from the disk are reprocessed and re-emitted at longer wavelengths \citep[e.g.][]{Guillochon:2013ApJ, Roth:2016ApJ}, from shocks caused by stream-stream collisions  \citep[e.g.][]{Piran:2015ApJ, Bonnerot:2017MNRAS}, or from some combination of both effects \citep{Lu:2020MNRAS}. To get a more complete picture of TDEs, we need a larger sample, and early detections during the phase in which the complicated debris geometry is still forming. Currently, we are discovering TDEs at a rate of $\sim10$ per year, mostly with wide-field optical surveys, such as, for example, the All-Sky Automated Survey for Supernovae (ASAS-SN; \citealt{Shappee:2013mna}), the Asteroid Terrestrial-impact Last Alert System (ATLAS; \citealt{Tonry:2018PASP}), and the Zwicky Transient Facility (ZTF; \citealt{Bellm:2019PASP}).

The upcoming optical Legacy Survey of Space and Time (LSST) at Vera C. Rubin Observatory (Rubin, \citealt{Ivezic:2019ApJ}) will be one of the most important projects in ground-based optical astronomy in the next decade. Its goal is to conduct a 10-year long survey to map $18,000$ deg$^2$ of the sky and observe billions of astronomical objects. One of the key scientific areas the LSST will explore is the observation of the transient sky. With its combination of a wide-field and deep imaging, relatively fast cadence (on order of $\sim 3$ days) and real-time data analysis, the LSST will provide a unique opportunity for discovering tens of thousands of new transients every night. 

We expect that the Rubin LSST will detect $\sim 1000$ TDEs per year \citep{vanVelzen:2011ApJ, Bricman:2020ApJ}. This will enable statistical studies of SMBH masses, accretion efficiency, TDE rates as a function of redshift and host galaxy type, etc. To make this possible, however, TDEs will need to be identified from the much larger sample of transients observed by Rubin (e.g. the supernova rate is $\sim1000$ times higher).

Some transients can act as TDE impostors \citep{Zabludoff2021SSRv}: they exhibit similar properties in their light curves, e.g. similar rise times or pre-peak optical colors (such as Type Ia SNe) or can be mistaken for a TDE due to their location in the host galaxy (mainly AGN flares).  Current classification methods for distinguishing between SNe, AGN and TDEs rely on spectroscopic observation, however, spectra will not be available for the majority of faint events detected by the Rubin LSST. Thus, a reliable photometric identification will be required for the majority of TDEs. 

In comparison to more common transient types, such as supernovae and active galactic nuclei, TDEs differ most markedly in their color evolution. They are bluer ($g-r<0,u-g<0$) and tend to remain blue for months after the peak \citep[e.g.][]{vanVelzen:2011ApJ, Hung:2017ApJ, vanVelzen:2020SSRv}. For photometric identification with Rubin, it will be essential that the observed light curves have sufficient multi-band coverage to allow color evolution measurements, especially in bluer bands (\emph{u, g, r}). We emphasize that observations in \emph{u} band are crucial to discern between SNe and TDEs - SNe populate the $u-g$ range between $0$ mag and $2$ mag, while TDEs are found to have mean $u-g$ color in the decay phase between $-0.5$ and $0$ \citep[]{vanVelzen:2011ApJ}.

The exact observing strategy of the Rubin LSST survey has not yet been selected. Some baseline properties of the observing strategy are determined, however, many parameters of the strategy are relatively open and can be tuned appropriately to maximize the scientific output of the survey. To satisfy the scientific requirements of the survey (see \citealt{Ivezic:2019ApJ} and \citealt{Bianco:2022ApJS} for details), LSST will continuously monitor an area of approximately $18,000$ deg$^2$ of the sky and each field in the sky will be visited about $800$ times in all six bands (\emph{u,g,r,i,z} and \emph{y}) together over the $10$ years of survey duration with a mean cadence between $3$ and $4$ days (in any of the filters). The Rubin LSST project has simulated a variety of proposed observing strategies, which address the modifications to this so-called baseline scanning law of the telescope. These strategies modify a variety of survey parameters, such as the footprint definition, the time devoted for observations in a certain filter, the exposure time per visit, the time difference between two subsequent observations of the same field in the sky, etc. The strategies are simulated using the Operations Simulator (OpSim; \citealt{Delgado:2014SPIE}) and its outputs are analyzed and visualized with the Metrics Analysis Framework (MAF; \citealt{Jones:2014}). 

As discussed in the opening of this focus issue \citep{Bianco:2022ApJS}, the observing strategy decision is largely inclusive of the community input. As members of the LSST Transient and Variable Stars Science Collaboration, we explore how different parameters of the survey strategy impact the observations of TDEs with Rubin and the possibility for their photometric identification based on the colors measured from the multi-band light curves. In this work, we use recent \footnote{At the time when most of the work for papers in this special issue was performed.} observing strategy simulations from Feature Based Scheduled (FBS) v1.5 and v1.7 survey strategy releases and develop new MAF-generated \texttt{TDEspop} metrics to evaluate the performance of different strategies for TDEs, and to assess which parameters of the observing strategy impact TDE observations most significantly. At this stage of the process towards the observing strategy decision, the metrics have to be relatively simple such that they can be executed efficiently for many different survey strategies. A full investigations into the efficiency and purity of machine learning methods for transient classification is not possible at this stage (e.g., this would require re-training the classifiers for each survey strategy). Such an investigation is currently on-going---but only for the baseline v2.1 cadence---in the ELAsTiCC\footnote{\url{https://portal.nersc.gov/cfs/lsst/DESC_TD_PUBLIC/ELASTICC/}} challenge.

This paper is organized as follows. In Section \ref{sec:Methods} we describe the methods we used to construct the \texttt{TDEspop} metrics. In Section \ref{sec:Outputs}, we present the outputs of the metric, discuss our results and highlight which parameters of the observing strategy impact the TDE observations with Rubin the most. Finally, we provide our conclusions in Section \ref{sec:Conclusion}.

\section{Methods}
\label{sec:Methods}

The survey strategy simulations in this cadence note are evaluated with the MAF \texttt{TDEspop} family of metrics (described in \ref{subsec:metrics}), which is available in the github repository in the \texttt{sims\_maf\_contrib}\footnote{\url{https://github.com/LSST-nonproject/sims_maf_contrib}}. The metrics take as an input a sample of light curves (at different redshifts) and the TDE rate.

\subsection{Input TDE light curves}

The \texttt{TDEspop} metric takes as an input a set of TDE light curves. We obtained the light curves of two known events, PS1-10jh \citep{Gezari:2012Nat} and iPTF-16fnl \citep{Blagorodnova:2017ApJ}. PS1-10jh is considered to be a normal TDE with typical temporal evolution with a rise e-folding time of $\approx 30$ d and the decay consistent with a power-law index of $\approx -1.5$. The characteristic decay time scale for PS1-10jh was $\approx 42$ d \citep{vanVelzen:2019ApJ}, and its peak absolute magnitude in \emph{g}-band $-19.6$ mag. iPTF16fnl, on the other hand, was the dimmest and the fastest evolving TDE observed so far, with a rise e-folding time of $\approx 10$ d, a decay consistent with a power-law index of $\approx -2.1$, and a characteristic decay timescale of $\approx 22$ d \citep{vanVelzen:2019ApJ}. Its peak absolute magnitude in \emph{g}-band was $-17.2$ mag. These two events are good representatives of a normal and a faster TDE (peak luminosities and decay rates for several TDEs were also reported in \citealt{Hinkle2020ApJ}).

\begin{figure*}[ht]
\begin{center}
\includegraphics[width=\textwidth]{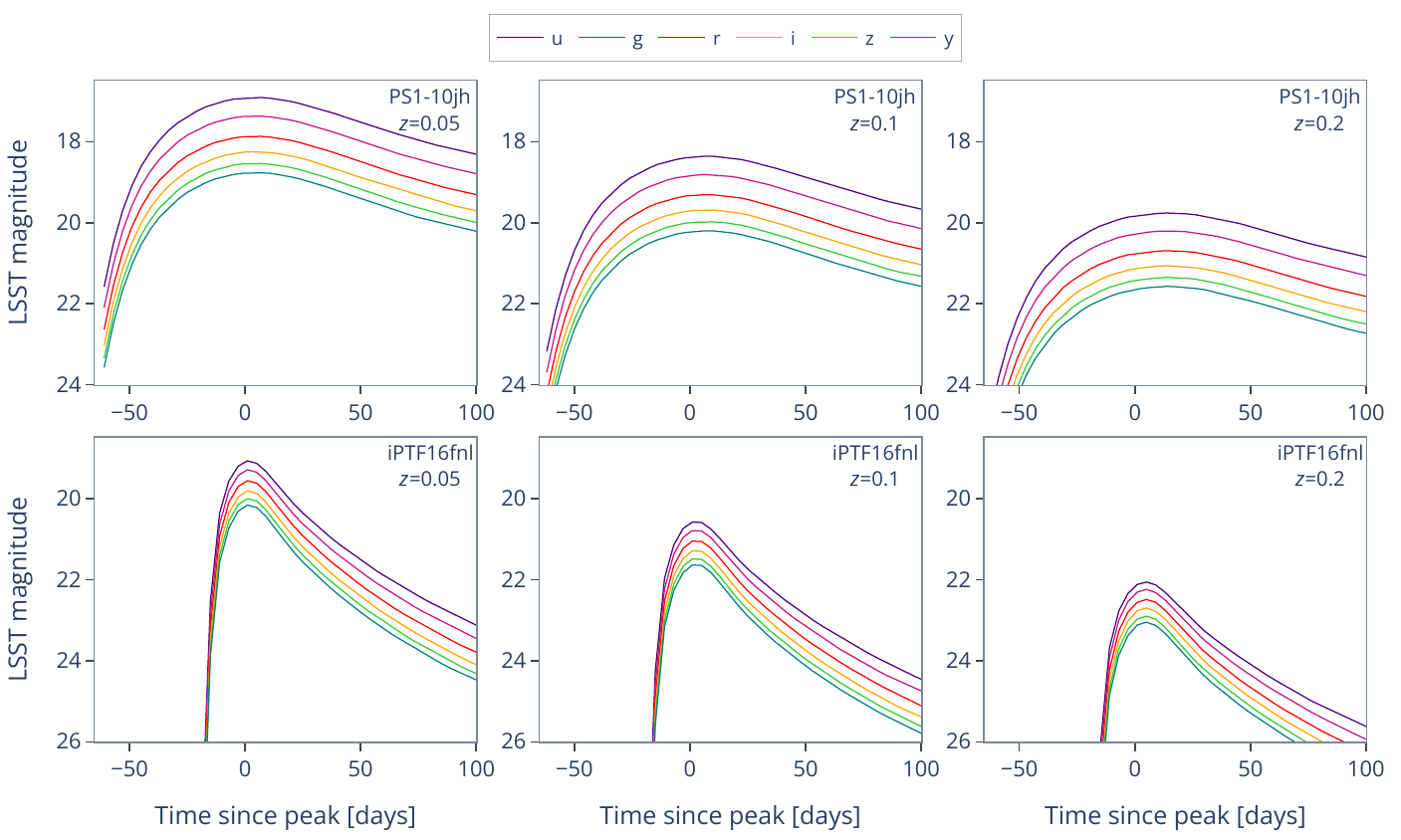}
\end{center}
\caption{Sample input TDE light curves for the \texttt{TDEs\_pop} metric. Rubin LSST magnitudes are plotted for PS1-10jh (top panels) and iPTF16fnl (bottom panels), at redshifts of $z=0.05$ (left), $z=0.1$ (center) and $z=0.2$ (right). The LSST single-epoch flux limit in the $r$-band is $\approx 24.5$. As such, each of these light curves can in principle be detected for a few months.
} 
\label{fig:metriclcs}
\end{figure*}

To produce continuously sampled light curves in Rubin LSST filters, we first fitted publicly available observational data of both events with the Modular Open Source Fitter for Transients (\texttt{MOSFiT}) \citep{Guillochon:2017bmg}. The fitting process returns a set of parameters of the event (e.g. black hole mass, impact parameter, etc.), which are given in Table \ref{table:FitPar} and are in line with results from \cite{Mockler:2018xne} and \cite{Nicholl:2020arxiv}. We then use \texttt{MOSFiT} again to generate mock light curves of both events in the LSST $ugrizy$ bands, using the parameters obtained in the previous step. The light curves were generated at three different redshifts, $z=0.05$, $z=0.1$ and $z=0.2$, and a K-correction was applied during the generation process. The continuously sampled light curves of both events in LSST filters are shown in Figure \ref{fig:metriclcs}. 

\begin{table*}[ht]
	\begin{center}
		\caption{The parameters obtained in MOSFiT ﬁts to PS1-10jh and iPTF16fnl light curves and used for generation of sample input TDE light curves.}
		\label{table:FitPar}
			\begin{tabular}{ l c c}  
				\hline 
				Parameter & PS1-10jh & iPTF16fnl \\ \hline
				Black hole mass $M_{\text{BH}}$ [$10^6 M_\odot$] & $17$ & $1.7$ \\
				Stellar mass $M_*$ [$M_\odot$] & $0.4$ & $0.1$ \\
				Scaled impact parameter $b$  & $0.997$ & $1$ \\
				Impact parameter $\beta$  & $0.899$ & $1.85$ \\
				Viscous delay $T_{\text{viscous}}$ [d] & $0.08$  & $0.04$  \\
				Efficiency $\eta$ & $0.09$ & $0.007$ \\
				Days since the first detection $t_\texttt{first fallback}$ [d] & $-12$ & $0$ \\
				Photosphere power-law exponent $l$ & $1.44$ & $1.7$ \\
				Normalization photospheric radius $R_{\text{ph,0}}$ & $6.3$ & $10.0$ \\
				\hline
			\end{tabular}
	\end{center}
\end{table*}

\subsection{TDE rates}

TDEs in our metric were distributed uniformly over the sky and over $10$ years of survey duration with a rate of $\approx 3$ TDEs per day. We also chose uniformly between both events (PS1-10jh and iPTF16fnl) and placed TDEs uniformly in all three redshift bins ($z=0.05$, $0.1$ and $0.2$). The assumed rate is conservative when compared to rates reported in \citet{Bricman:2020ApJ}, where the authors assumed the TDE rate per galaxy per year is $10^{-5}$ (as shown in \citealt{vanVelzen:2011ApJ} and \citealt{Holoien:2015pza}) and found that between $10$ and $20$ TDEs could be observed per day. Furthermore, the TDE rate assumed in this work is even more conservative, since recent estimates show that, after accounting for the steep slope of the TDE optical luminosity function, the TDE rate is actually closer to $10^{-4}$ galaxy$^{-1}$ year$^{-1}$ \citep[e.g.][]{vanVelzen:2018ApJ, Hung:2017ApJ}. The sources with a low optical luminosity (such as iPTF16fnl) dominate the volumetric rate, while for a flux-limited sample (such as detected in LSST), the observed rate as a function of luminosity is roughly constant \citep{vanVelzen:2018ApJ}. However, our goal here is not to accurately predict the absolute number of TDEs observed by Rubin, but to evaluate the scientific performance of Rubin for TDE studies, including the quality of the light curves for identifying and understanding the detected TDEs. The output of the metric is in any case the fraction of injected TDEs, which pass our criteria (defined in \ref{subsec:metrics}), and this fraction is not influenced by the order of magnitude difference between the per galaxy rates reported above.

\subsection{Metrics}
\label{subsec:metrics}
TDE observations with Rubin LSST are synthesized with MAF for a given observing strategy and the simulated light curves are analyzed to determine what fraction of events meets each of three identification criteria. The criteria on the number of detections in a certain time interval of the light curve listed here represent different levels of scientific utility of Rubin data. In each of them we assume that a TDE is identified if the following minimum requirements are met:
\begin{itemize}
    \item \texttt{prepeak}: there are at least two detections before the peak of the light curve;
    \item \texttt{some\_color}: there is at least one detection pre-peak ($t < t_{\text{peak}} - 10$ days), at least $3$ detections in at least $3$ different filters within $10$ days of the peak, and at least $2$ detections in at least $2$ different filters between $10$ and $30$ days after the peak;
    \item \texttt{some\_color\_pu}: there is at least one detection pre-peak ($t < t_{\text{peak}} - 10$ days), at least one detection in \emph{u}-band and at least one detection in another filter within $10$ days of the peak, and at least one detection in \emph{u}-band and at least one detection in another filter between $10$ and $30$ days after the peak.
\end{itemize}

The \texttt{prepeak} metric simply reflects on how many TDEs are observed before the peak. Detecting events pre-peak is crucial for fitting light curve models to derive physical quantities. However, light curves passing only this metric (and not the other two metrics) are typically sampled quite sparsely (see the upper left panel of Figure \ref{fig:recoveredlcs}) and are not particularly useful for photometric identification. Nevertheless, we include it here for illustrative purposes.

Light curves passing the \texttt{some\_color} and \texttt{some\_color\_pu} metric requirements should, on the other hand, be more suitable for photometric identification of TDEs. The color metrics defined above require a data point at $t < t_{\text{peak}} - 10$ days, which is important for accurately determining the peak time of the light curve, the peak magnitude, and the rise to peak timescale. The requirement to determine the peak time to within 10 days is the bare minimum for a reliable estimate of the black hole mass.
From Kepler's law, the most bound debris returns to pericenter on a timescale $t = 1/\sqrt{2} \pi (G M_* / R_*^3)^{-1/2} (M_{\rm BH}/M_*)^{1/2}$, which to first order determines the rise time of the light curve. Thus, for a typical black hole of $10^6$\,M$_\odot$, an uncertainty of $10$ days in peak time corresponds to an uncertainty in mass of $5\times10^5$\,M$_\odot$, i.e. $50\%$. Detections between $10-30$ days after peak will allow a measurement of the rate of fading, a proxy for the fallback rate that also encodes the mass of the star and impact parameter \citep{Guillochon:2013ApJ}.

The additional requirements in the \texttt{some\_color} metric allow for sufficient measurement of color evolution of the event during the early phase of the light curve (e.g. within $\sim 1$ month after the first observation). TDEs show almost no change in $g-r$ color during this phase, whereas SNe redden by $\approx0.01-0.04$ mag per day (see \citealt{vanVelzen:2021ApJ} Fig. 1). Although studied mainly in $g-r$ to date, the lack of color evolution in TDEs is true for arbitrary bands, as it relates to the lack of photospheric temperature evolution (SNe cool down, TDEs do not).
A transient with apparent magnitude $m\approx24$ (23) is detected by Rubin with an uncertainty of $\sim0.04$ (0.02) mag in each band. If the color is measured between observations $\sim5$ days apart, a change in color of $g-r\sim0.02$ (average for a SN, discriminating it from a TDE) is detected with a signal-to-noise ratio of $\approx 2$ (4).

The most relevant set of requirements are encompassed in the \texttt{some\_color\_pu} metric, which is in addition to measuring the color evolution and event's properties also sensitive to \emph{u}-band observations. As discussed earlier, these will be crucial for photometric recognition of TDEs, since a sufficient number of observations in bluer bands should allow for a TDE identification based on Rubin photometric data only. A caveat of this metric, however, is that it is very sensitive to detections in $u$-band at certain and limited intervals of time after the peak of the light curve. Due to this requirement, some bright events might not pass the metric, although we could expect they would be recognized otherwise simply due to their brightness and longer duration. 

For each set of identification requirements, the output metric values (i.e. the fraction of identified TDEs) is calculated for each observing strategy and for each field in the sky. The median over all metric values (i.e. over all visited fields in the sky) serves as a parameter for evaluating the performance of a certain observing strategy. The higher its value, the better a particular survey strategy is for identifying TDEs. 

There are two additional metrics we constructed, the so called \texttt{TDEQuality} metrics, which measure the light curve quality of identified TDEs. To each TDEs passing the \texttt{some\_color} or \texttt{some\_color\_pu} requirements, separately, we assign a ``score'', accounting for how well a light curve is sampled. If the TDE does not meet the color requirements set by our two original metrics, the score is automatically $0$. If it does pass the color metric, then we count the number of detected data points in the light curve. The number of data points is then divided by a user defined time-interval, in which it is specifically important for parameter-extraction, model-fitting and interpolation purposes that the light curves are covered well. By default, this time interval is defined as $-30$ days $ < t_{\text{peak}} < 100$ days (which includes most detection baseline above the flux limit, see Fig.~1). Since the LSST's design by default takes visits in a pair to the same given field on the sky in the same night, this means that if the light curve quality score is equal to $1$, then the TDE is, on average, observed every second night. Lower scores mean that the cadence is more sparse, and higher scores that the TDE is observed more often than every other night, e.g. a TDE withing a Deep Drilling Field (Deep Drilling Fields - DDFs, are observed more often than every other night) can have a score higher than $1$. The scores of all TDEs that passed the color requirements are then averaged over the whole survey footprint and survey duration, resulting in one number, which is an estimate for how well the light curves are covered for a certain observing strategy. Most of the resulting values for different observing strategies lie between 0 and 1. Individual TDEs with light curve quality scores higher than $1$ are rare and do not affect significantly the final average over all TDEs.

\section{Results and discussion}
\label{sec:Outputs}
The \texttt{TDEspop} metrics can, in general, generate three outputs: i) the recovered light curves of events simulated with a certain observing strategy; ii) the fraction of input TDEs that satisfied the metric requirements; and iii) the light curve quality score averaged over TDEs that passed one of the color metrics. In the following sections we present our results on the metric outputs and comment on which survey strategy parameters impact the Rubin TDEs performance most. Particular emphasizes are given to the changes of survey strategy parameters, which were presented in the Cadence Notes (see \citealt{Bianco:2022ApJS} for more details).

\subsection{Recovered light curves}
\begin{figure*}[ht]
\begin{center}
\includegraphics[width=\textwidth]{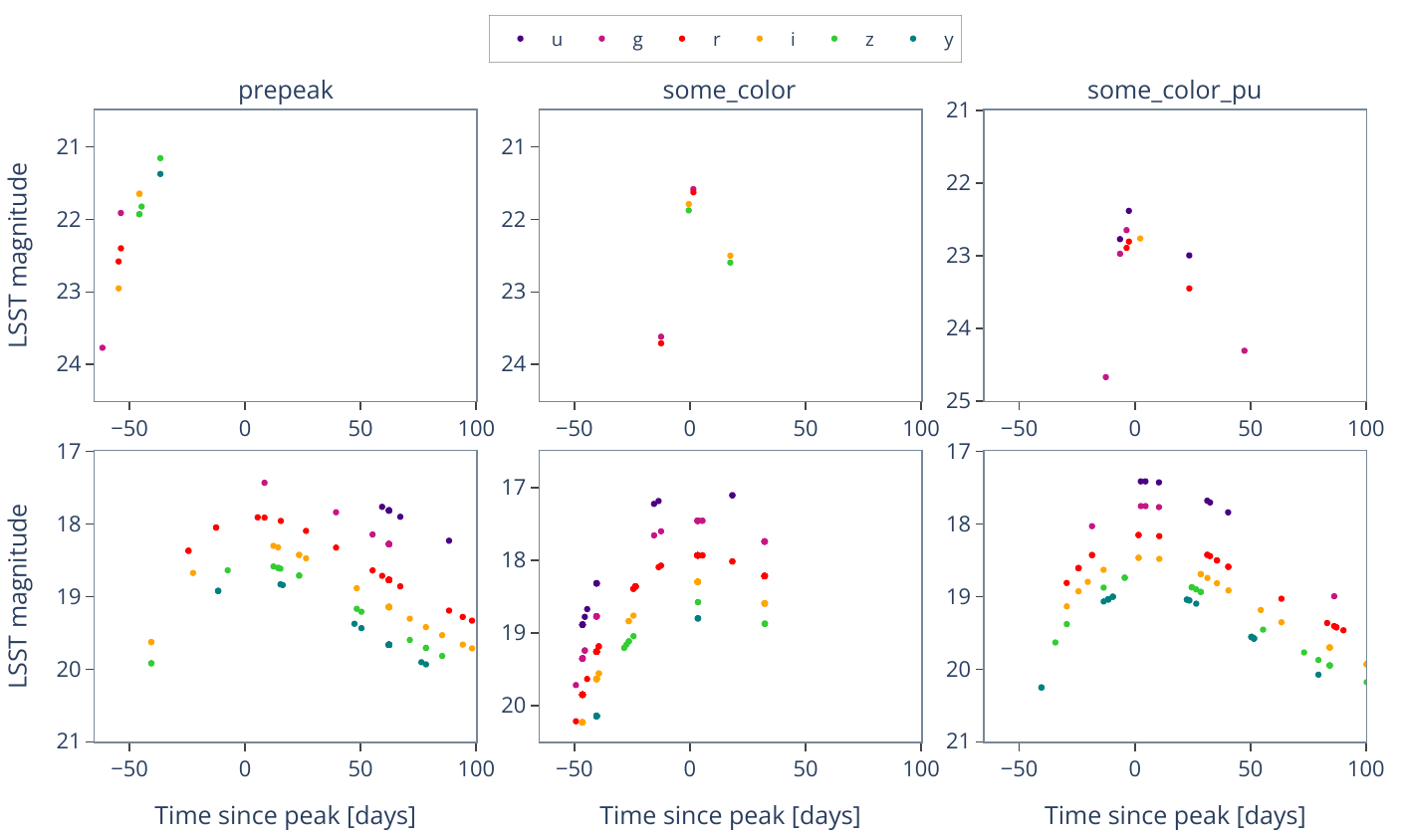}
\end{center}
\caption{Recovered TDE light curves simulated with the observing strategy \texttt{baseline\_nexp2\_v1.7\_10yrs}. Each column shows two example of light curves, which passed the metric requirements of \texttt{prepeak} (left), \texttt{some\_color} (center) and \texttt{some\_color\_pu} (right) metrics. The upper row shows the ``worst'' light curve of all recovered light curves, with the lowest light curve quality score (i.e. the number of data points in a time interval between $30$ days before and $100$ days after the peak), while the bottom row shows the ``best'' example with the highest light curve quality score. For example, the light curve quality scores of the "bad" light curve detected with the \texttt{some\_color\_pu} metric is $0.085$, with an average time interval between two observations of $23$ days, while the light curve quality score of the "good" TDE detected with the \texttt{some\_color\_pu} metrics is $0.67$, with an average time between two subsequent observations of $3$ days.} 
\label{fig:recoveredlcs}
\end{figure*}

Figure \ref{fig:recoveredlcs} shows $6$ examples of recovered TDE light curves, simulated with the baseline observing strategy \texttt{baseline\_nexp2\_v1.7\_10yrs}. We show two light curves for each of the \texttt{TDEspop} metrics: \texttt{prepeak}, \texttt{some\_color} and \texttt{some\_color\_pu}. The light curve in the upper panel is an example of a ``worst'' candidate, and the light curve in the bottom panel represents the ``best'' TDE. The majority of light curves passing the \texttt{prepeak} metric are sparsely sampled and close to the worst-case light curve, while \texttt{prepeak} TDEs that are sampled more frequently usually also pass (one of) the color metrics.

From Figure \ref{fig:recoveredlcs} it is important to note that a TDE passing the basic metric requirements listed above still might not be sufficient for identification due to sparse sampling of the light curve. However, with the bare minimum requirements, we should be able to measure the color of the transient. The color information could be very useful in making decisions about triggering follow-up observations. 

To fully exploit the Rubin data, on the other hand, light curves such as those in the bottom panels of Figure \ref{fig:recoveredlcs} are far more useful. Not only would these events be photometrically classified earlier, they are also very well sampled, which is important for model-fitting approach of extracting parameters. 

\subsection{Fraction of identified TDEs by TDE type and redshift}

\begin{figure}[ht]
\begin{center}
\includegraphics[width=.9\linewidth]{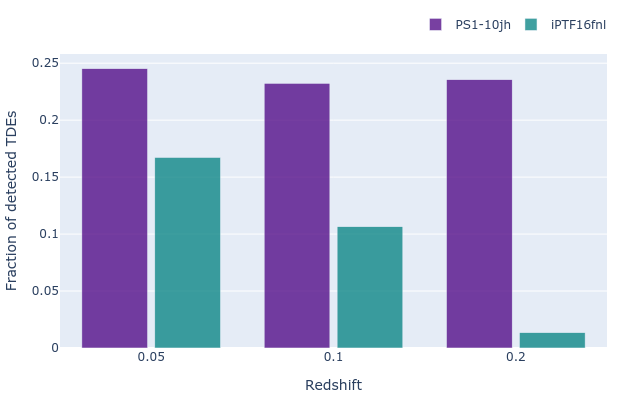}
\end{center} 
\caption{Fraction of TDEs passing the \texttt{some\_color} metrics in the \texttt{baseline\_nexp2\_v1.7\_10yrs} observing strategy by their type (``fast and faint'' TDE or iPTF16fnl and ``normal'' TDE or PS1-10jh) and redshift.}
\label{fig:tdesbyredshift}
\end{figure}

Our input light curves represent two TDEs types: a ``fast and faint'' (iPTF16fnl) and a ``normal'' TDE (PS1-10jh), placed at three different redshifts. In Figure \ref{fig:tdesbyredshift} we show the fraction of TDEs which passed the \texttt{some\_color} metric in the ``main survey'' (Wide–Fast–Deep or WFD fields) of the \texttt{baseline\_nexp2\_v1.7\_10yrs} observing strategy (i.e. TDEs ``detected'' by this metric) by their type and redshift. We show the fraction of detected TDEs compared to all detected TDEs (both types, all redshifts), since the absolute numbers are less useful for comparison purposes with other metrics.

The ``normal'' TDE type (PS1-10jh) has approximately the same number of detected candidates at each redshift applied. The ``fast and faint'' TDE type (iPTF16fnl) has the highest number of detected candidates at the closest redshift, and their number rapidly decreases with larger redshift. This implies that normal events are expected to be more easily identified at all three redshifts, while the majority of the sample of the detected faster events will be made of events at lower redshifts. The distribution shown in Figure \ref{fig:tdesbyredshift} looks similar if instead of \texttt{some\_color} metric the \texttt{some\_color\_pu} metric is used.

\subsection{Fraction of identified TDEs and quality of their light curves}

\begin{figure*}[ht]
\begin{center}
\includegraphics[width=.85\textwidth]{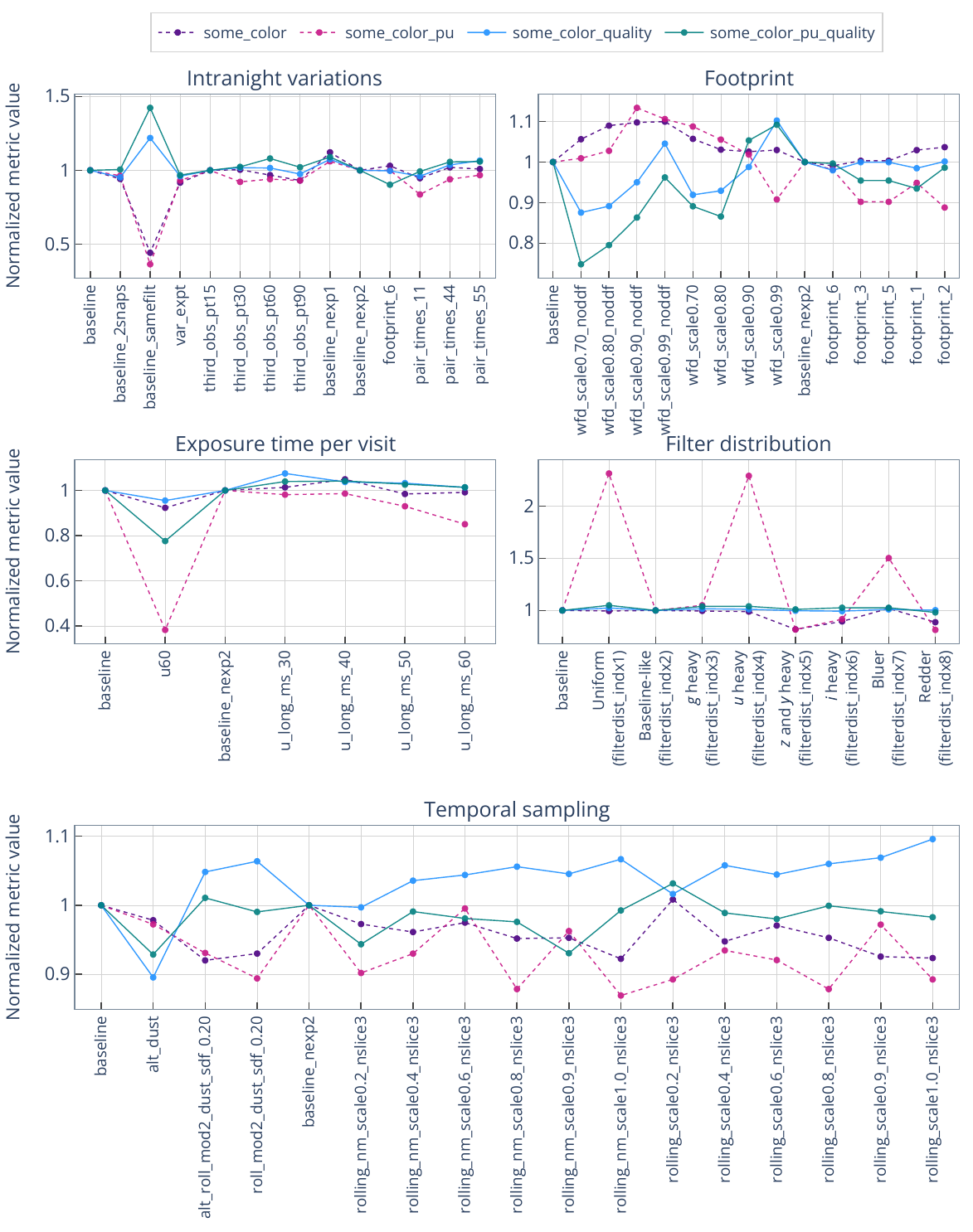}
\end{center}
\caption{Metric output values or the fraction of TDEs (out of injected ones) passing the metric requirements for two colour \texttt{TDEspop} metrics: \texttt{some\_color} (purple), \texttt{some\_color\_pu} (pink), and the quality metric output for the \texttt{some\_color\_quality} (blue) and \texttt{some\_color\_pu\_quality} (green) metrics, are shown for a variety of observing strategies. We show results for survey strategy simulations, which are part of FBS v1.5 (\texttt{baseline}, \texttt{var\_expt}, \texttt{third\_obs}, \texttt{wfd\_scale}, \texttt{footprint}, \texttt{u60}, \texttt{alt}, \texttt{roll\_mod2}) and FBS v1.7 (\texttt{baseline\_nexp}, \texttt{pair\_times}, \texttt{footprint1-8}, \texttt{u\_long}, \texttt{rolling}) releases. The metric outputs are normalized to the baseline observing strategy of the corresponding survey strategy release. The lines connecting the values of each metric are shown for representative purposes only and do not imply any trend from one strategy to another.}
\label{fig:metricvalues}
\end{figure*}

Figure \ref{fig:metricvalues} shows the metric outputs, or the fraction of input TDE that satisfied the metric requirements, for the \texttt{TDEspop} color metrics \texttt{some\_color} and \texttt{some\_color\_pu}. We do not include results for the \texttt{prepeak} metric, since as mentioned above, light curves passing this metric are not particularly useful for photometric identification. We also show results for quality metrics \texttt{some\_color\_quality} and \texttt{some\_color\_pu\_quality}. The metric values are shown for a variety of observing strategies, in which different survey parameters are changed with respect to the baseline. We sample several strategies from survey strategy simulation releases of FBS v1.5 and FBS v1.7. For comparison purposes we chose a subset of survey strategies presented in \cite{Jones2020}. More information about each individual survey strategy shown in Figure \ref{fig:metricvalues} can be found in Appendix \ref{appendix:a}.

The metric values are normalized to the corresponding baseline strategy of the same simulation release, in order to asses how changing a certain strategy parameter impacts the fraction of detected TDEs and the quality of light curves with respect to the baseline strategy. 

The fraction of TDEs satisfying each of the metric requirements can also be linked to the number of TDEs satisfying certain metric requirements per year (since the input is always $1000$ per year), while the light curve quality score is related to the mean time interval between two subsequent observations in the light curve (or the inter-night cadence). For the baseline observing strategy we report the metric values together with the fraction of TDEs passing each metric requirements, the light curve quality scores and the inter-night cadence in Table \ref{table:TDEnumbers}. 

\begin{table*}[ht]
	\begin{center}
		\caption{The metric values (i.e. the fraction of TDEs passing the metric requirements) of both color metrics (\texttt{some\_color} and \texttt{some\_color\_pu}), the light curve quality scores (i.e. the \texttt{some\_color\_quality} and \texttt{some\_color\_pu\_quality} metric values) and the average inter-night cadence for identified TDEs with both color metrics are reported for the \texttt{baseline} observing strategy.
		}
		\label{table:TDEnumbers}
		\begin{threeparttable}
			\begin{tabular}{ l c}  
			\hline 
			Metric & \texttt{baseline} observing strategy\tnote{a} \\ \hline
            \texttt{some\_color} metric value &  $0.075$  \\
            \texttt{some\_color\_quality} metric value &  $0.37$  \\
            Average inter-night cadence for \texttt{some\_color} TDEs & $5.4$ days \\ \hline
            \texttt{some\_color\_pu} metric value &  $0.015$  \\
            \texttt{some\_color\_pu\_quality} metric value &  $0.4$  \\
            Average inter-night cadence for \texttt{some\_color\_pu} TDEs & $5$ days \\ \hline            
			\end{tabular}
		  \begin{tablenotes}
		  	\footnotesize{
			\item[a] The FBS v1.5 baseline strategy is named \texttt{baseline} in Figure \ref{fig:metricvalues} and the FBS v1.7 baseline strategy is named \texttt{baseline\_nexp2} in Figure \ref{fig:metricvalues}. The metric values for both baselines are similar.
		  	}
		\end{tablenotes}
	\end{threeparttable}
	\end{center}
\end{table*}

\subsubsection{Intra-night cadence variations}
\label{subsec:intranight}

The upper left panel of Figure \ref{fig:metricvalues} shows results for simulations varying the intra-night cadence. By default, two visits are obtained per night per field in two different filters separated by $22$ minutes. The intra-night variations address the impact of either having a visit composed of $2\times15$ s exposure or $1\times30$ s (\texttt{baseline, baseline\_2snaps, baseline\_nexp1, baseline\_nexp2}), performing both visits to a given field in the sky in the same filter (\texttt{baseline\_samefilt}), adding an additional visit to the field during the night (\texttt{third\_obs}) or separating the two visits per field by different time intervals (from $11$ to $55$ min in \texttt{pair\_times}). 

We find that the effect of adding the third visit or separating the two visits for more/less than $22$ minutes generally reduces the efficiency of TDE detection. The baselines with $2\times15$ s exposures are generally performing worse, since the total number of visits per field is reduced when compared to the $1\times30$ s baseline strategies (due to $2$ s read-out time between two exposures). Performing both visits to a given field in the sky in the same filter severely decreases the metric values of both color TDE metrics. The cadence of the \texttt{baseline\_samefilt} strategy is too low to provide two pairs of visits in $10$ days around the peak, in different filters, which severely reduces our ability to measure the color of the event in the most important part of the light curve. 

Here we would also like to make a point that the pairs of visits in the same night are not strictly necessary for light curve classification for "slow" transients such as TDEs. While it is useful to measure a color from a pair of observations in different filters in a single night, it is also possible to measure the color from a model that fits the light curve in those particular filters as a function of time. Another possibility is to train machine learning classification to work with non-simultaneous flux measurements in different filters. Strict pairs are therefore not necessary, but observations in different bands do need to be close in time to avoid excessive extrapolation. For TDEs, getting two observations in different filters in $2$ to $5$ nights would be acceptable. 
We showed in section \ref{subsec:metrics} that observations within $\approx5$ days is sufficient to measure the color evolution that separates SNe from TDEs. However, if observations in different filters are too widely separated in time, evolution in luminosity can be confused for evolution in color. One can mitigate this by fitting models to the light curve, though we then need to worry about extrapolation errors. If observations in different filters are separated by $N$ days, an extrapolation error of $\approx0.01$ mag per day becomes a magnitude error of $0.01\sqrt{N}$, which exceeds the average colour evolution per day of a SN after $\approx 5$ days.
With such relatively loose limits on the time range between two observations in different filters our metric allows for more flexibility than many other metrics, which require visits in different filters to be taken within the same night. 

The light curve quality is similarly not affected much by the intra-night changes. The peak of the light curve quality score for the \texttt{baseline\_samefilt} is due to a smaller overall number of TDEs detected: among them there is a larger fraction of TDEs in DDFs (which are visited more often and give better sampled light curves) which leads to a higher score ($>1$). However, the coverage of TDEs outside DDFs remains approximately the same.

\subsubsection{Survey footprint variations}

The upper right panel of Figure \ref{fig:metricvalues} shows results for simulations varying the footprint of the survey. Typically, the survey footprint of the baseline strategies is $\approx 18,000$ deg$^2$, where $90$\% of time is devoted to Wide–
Fast–Deep (WFD) observations. The \texttt{wfd\_scale} family devotes from $70$\% to $99$\% of observing time for the main WFD survey, while the rest is devoted to mini-surveys. If DDFs are included, they are allocated $5$\% of the available survey time. The \texttt{footprint} family of strategies explore the effects of extending the WFD area beyond baseline strategy footprint (simulations \texttt{footprint\_1}, \texttt{footprint\_3}, \texttt{footprint\_5} and \texttt{footprint\_6}) including high dust-extinction regions. The number of visits to a given field in the sky drops by $\approx10$\% in these simulations. Strategy \texttt{footprint\_2} has a reduced WFD area with respect to baseline. 

The metric values of our color metrics are typically slightly higher when compared to the baseline, if more time is devoted to WFD observations and there are no DDFs observed, since the total number of visits to a given field in the sky is increased. The effect of a larger WFD is not reflected in our TDE color metrics, since they only measure the fraction of TDEs that satisfy the minimum identification requirements.

The changes in the light curve quality are more apparent, if the footprint or time devoted to WFD observations is changed. The light curve quality increases, if more time is devoted to WFD observations. The light curve quality score decreases, if the WFD area is increased, and has a similar value to the baseline value, if the WFD area is reduced.

We expect that a smaller footprint would still be more beneficial, since it would provide denser sampling of the light curves, which would enable observations of structures in the light curves (that might be missed otherwise), discoveries of fast-evolving TDEs and a more accurate determination of their physical parameters (such as black hole mass, stellar mass, etc.).

According to estimates of the current system performance there might be additional observing time (as much as $10$\% of the survey time) available for visits of mini-survey and DDFs or other scientific use. We stress that more time for DDFs or mini-surveys would not necessarily benefit transient science, unless the mini-surveys are dedicated especially to transient hunt, such as, for example, the multi-messenger Target of Opportunity (ToO) observations. The best use of this extra $10$\% of time is to skew the filter weights more towards the blue end. As shown in Figure 19 of \cite{Jones2020}, reproduced below (Figure \ref{fig:metriccompare}), there is an enormous increased return for transient science by adding more weight to the blue end. This is most clearly seen in the TDE metric, but the SN Ia metric also shows a positive response, albeit with smaller amplitude (this could reflect the fact that this metric is optimized for detections, but not for photometric typing). Finally, we note that based on the investigation of photometric typing efficiency by \cite{Villar:2018ApJ}, we can be certain that bluer weighting also helps in identifying -- and, crucially, inferring physical parameters for -- transients that are not included in the current metrics, in particular superluminous SNe. This is also likely true of the emerging class of `rapidly evolving' or `fast, blue' transients \citep{Drout:2014ApJ, Margutti:2019ApJ}.

\subsubsection{u-band exposure time}

Since FBS v1.7, the visit in the baseline observing strategy is defined as $2\times15$ s exposures (in previous simulation runs, the default visit consisted of $1\times30$ s exposure). $2\times15$ s exposures are more efficient in detecting cosmic rays and rapid variability of objects and will be adopted by the survey at least in the commissioning phase. A possible exception for this visit definition is the $u$-band, for which the combination of the camera readout noise and the low sky background argues for longer exposures. Changing the $u$-band exposure time to $1\times50$ s exposure would result in $0.5$ mag deeper limiting magnitudes, which would improve the photometric redshift and photometric metallicity measurements, as well as transient classification. 

The center left panel of Figure \ref{fig:metricvalues} shows results for varying the $u$-band exposure time due to readout noise. The \texttt{baseline} assumes $1\times30$ s exposure in $u$-band, while the \texttt{baseline\_nexp2} assumes $2\times15$ s exposures. The \texttt{u\_long} simulations increase the exposure time in $u$-band from $1\times 30$ s to $1\times 60$ sec, while retaining a similar number of visits to a field in $u$-band. The \texttt{u60} simulations has a $1\times60$ s exposure, but cuts the number of $u$-band observations in half. 

The effects of changing the $u$-band exposure time on the \texttt{some\_color} metric are minor, they become more obvious in the case of \texttt{some\_color\_pu} metric, which is by definition very sensitive to the number of visits in $u$-band. Because for longer $u$-band exposure times, the overall number of visits in $u$-band is reduced, this affects the \texttt{some\_color\_pu} metric values. The metric values are particularly low for \texttt{u60} simulation due to a significantly lower number of total $u$-band visits.

The light curve quality is not particularly affected by the $u$-band exposure time. The minimum for the \texttt{some\_color\_\_pu\_quality} is not attributed to the light curve coverage, but rather again to a small sample of TDEs that passed the \texttt{some\_color\_pu} metric requirements. 

We stress here that in order to reduce the effect of the $u$-band readout noise, it is beneficial to increase the $u$-band exposure time, however, for transient identification purposes the number of $u$-band visits should be kept the same.

\subsubsection{Filter distribution among visits}

The center right panel of Figure \ref{fig:metricvalues} shows the results for changing the filter distribution among visits. In the baseline observing strategies the following fractions of observing time are allocated per band: $0.07$, $0.10$, $0.22$, $0.22$, $0.19$, $0.19$ in $u$, $g$, $r$, $i$, $z$ and $y$, respectively \citep{lsstSRD}. The \texttt{filterdist\_v1.5} simulations vary the distribution of filter weights (i.e. the observing time allocated per band). The fractions of observing times allocated per filter for these simulations is shown in Table \ref{table:filterdistrib}. The \texttt{filterdist} family of survey strategy simulations has a different footprint than the baseline, therefore the metric outputs shown in Figure \ref{fig:metricvalues} are normalized against \texttt{filterdist\_indx2} (not against the baseline), which has a similar filter distribution as the baseline strategy. 

\begin{table*}[ht]
\centering
\caption{Fraction of observing time allocated per filter in \texttt{filterdist} (\texttt{indx1-8}) survey strategy simulations.}
\label{table:filterdistrib} 
\begin{tabular}{ l c c c c c c}  
\hline 
Strategy & $u$ & $g$ & $r$ & $i$ & $z$ & $y$ \\ 
\hline
Uniform (\texttt{indx1}) 						& 0.16 & 0.16 & 0.16 & 0.17 & 0.17 & 0.18 \\ 
Baseline-like (\texttt{indx2})					& 0.06 & 0.09 & 0.22 & 0.22 & 0.20 & 0.20 \\
\emph{g} heavy (\texttt{indx3})				& 0.06 & 0.19 & 0.20 & 0.19 & 0.18 & 0.18 \\
\emph{u} heavy (\texttt{indx4}) 				& 0.16 & 0.09 & 0.19 & 0.19 & 0.18 & 0.18 \\
\emph{z} and \emph{y} heavy (\texttt{indx5}) 	& 0.05 & 0.07 & 0.17 & 0.17 & 0.26 & 0.26 \\
\emph{i} heavy (\texttt{indx6}) 				& 0.06 & 0.08 & 0.20 & 0.29 & 0.18 & 0.18 \\
Bluer (\texttt{indx7}) 						& 0.09 & 0.12 & 0.20 & 0.20 & 0.18 & 0.19 \\
Redder (\texttt{indx8}) 						& 0.06 & 0.08 & 0.20 & 0.22 & 0.21 & 0.22 \\
\hline
\end{tabular}
\end{table*}

\begin{figure}[ht]
\begin{center}
\includegraphics[width=.9\linewidth]{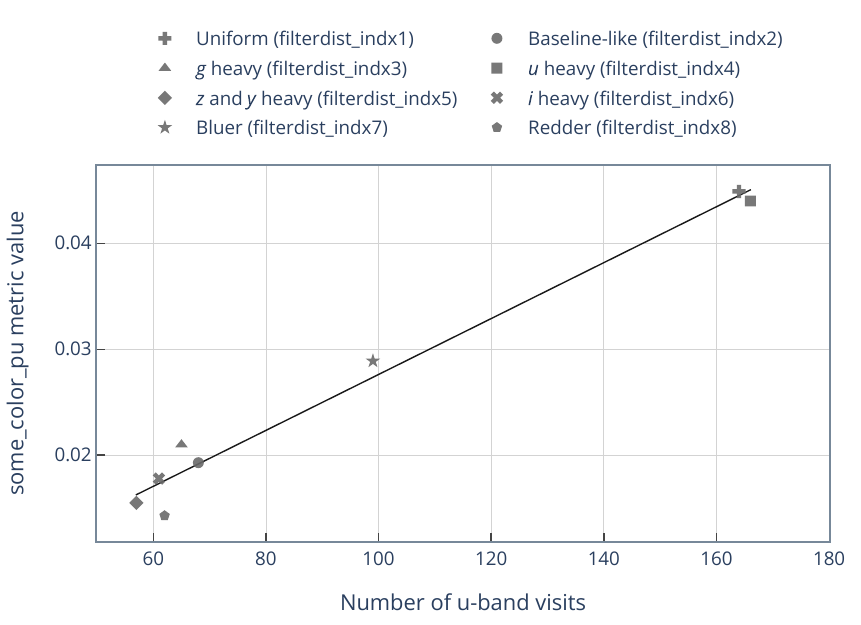}
\end{center} 
\caption{\texttt{some\_color\_pu} metric value as a function of number of \emph{u}-band visits in \texttt{filterdist} family of survey strategy simulations. The metric value scales linearly with the number of observations in \emph{u}-band.}
\label{fig:ubandscaling}
\end{figure}

We find that the \texttt{some\_color} metric values slightly decrease for strategies with more observing time in redder bands (i.e. the strategies named $z$ and $y$ heavy, $i$ heavy and redder). The change in the \texttt{some\_color\_pu} metric, however, is very substantial. Strategies with heavier weights on $u$-band observations perform significantly better than those with more observations in redder bands. For example, the uniform, $u$ heavy and bluer simulations (or \texttt{filterdist\_indx1}, \texttt{filterdist\_indx4} and \texttt{filterdist\_indx7}, respectively) yield $1.5$ to $2.3$ times larger number of identified TDEs with the \texttt{some\_color\_pu} metric solely by allowing for more observations in $u$-band. The metric values scale linearly with the number of observations in $u$-band (see Figure \ref{fig:ubandscaling}).

\begin{figure*}[ht]
\begin{center}
\includegraphics[width=.85\textwidth]{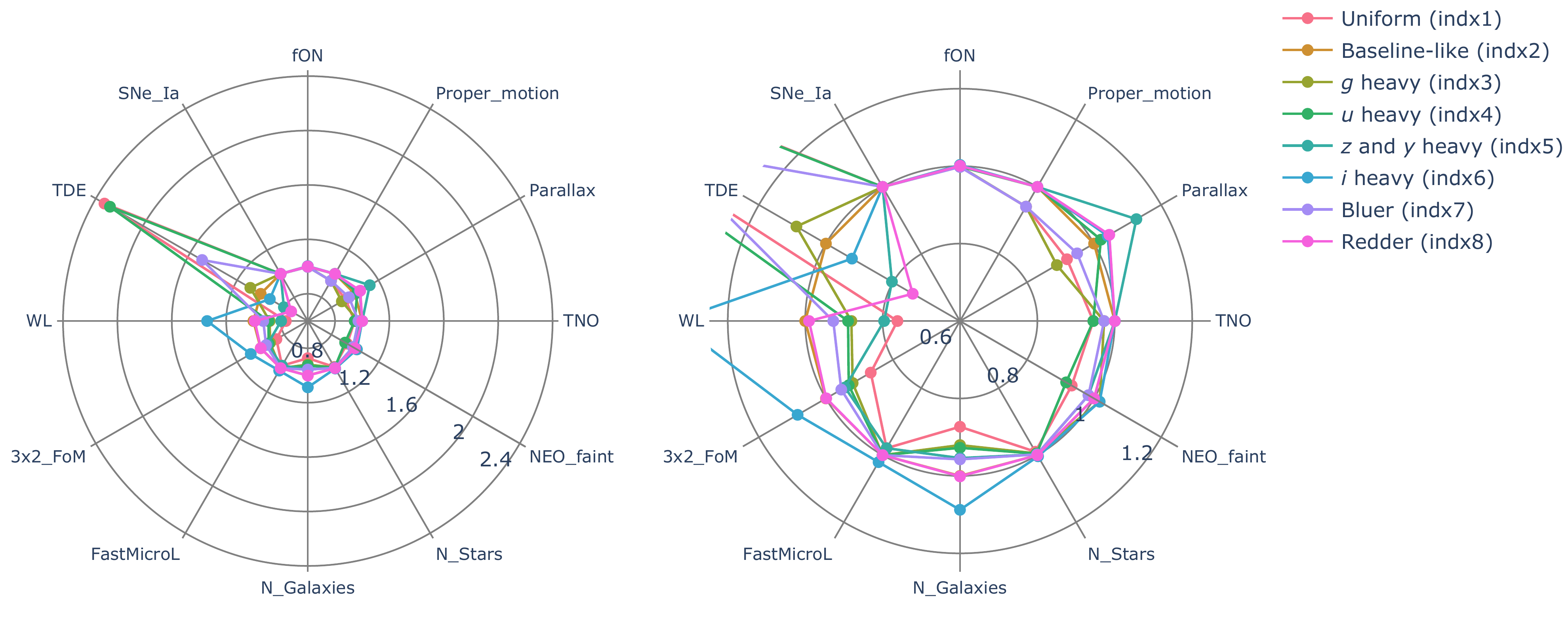}
\end{center} 
\caption{Metric values for 12 science cases as a function of \texttt{filterdist} family simulations. In the left panel we plot the metric values normalized to \texttt{filterdist\_indx2}, to show the significant effect of a larger number of \emph{u}-band observations on the TDE \texttt{some\_color\_pu} metric. In the right plot we show the same values, but zoomed to make variations between metrics more obvious (similar to Figure 19 in \citealt{Jones2020}). We see a very large increase of detection efficiencty for TDEs (and other blue transients), compared to a much smaller change for metrics of the other science cases.}
\label{fig:metriccompare}
\end{figure*}

From these simulations we expect that observing strategies with more $u$-band observations are significantly better for photometric identification of TDEs. We also investigated how a larger number of $u$-band observations influences other science cases. A set of the most important science-related metrics was described in \cite{Jones2020}. We chose $12$ of those metrics, which evaluate observing strategies for different science cases: the median number of visits to a given field (fON), the median proper motion error (proper motion), the median parallax error (parallax), Trans-Neptunian Object (TNO) population completeness due to discovery with the LSST after $10$ years (TNO), faint Near Earth Object (NEO) population completeness due to discovery with the LSST after 10 years (NEO\_faint), number of stars detected at the $5\sigma$ level in $i$-band (N\_stars), number of galaxies observed across the entire survey footprint (N\_galaxies), the discovery rate of fast microlensing events (FastMicroL), $3\times$ figure of merit for weak lensing and large scale structures, which measures shear-shear, galaxy-shear and galaxy-galaxy correlations ($3\times2$ FoM), mean number of visits per pointing across extragalactic footprint to estimate weak lensing systematics (WL), the \texttt{some\_color\_pu} TDE metric (TDE) and supernovae Ia metric, which measures how well the SNe alerts, that can act as follow-up triggers, are produced (SNe\_Ia). We obtained the metric values for all $12$ cases from the publicly available database of all observing strategies\footnote{\url{http://astro-lsst-01.astro.washington.edu:8081/}} and the corresponding metric outputs. By comparing these metric values together with the results for TDEs, we could asses the impact of filter distributions on each science case. We show these results in Figure \ref{fig:metriccompare}.

While the TDE metric values peak for bluer simulations, we see that strategies with a larger number of $u$-band observations perform well for other science cases as well. However, if the strategies prefer more observations in redder filters, the TDE metric values are considerably lower in comparison with other science cases. This is where we can make big gains. Bluer weighting, more \emph{u}-band or \emph{g}-band heavy, is very helpful for TDE science. But stronger \emph{g}-band weighting is better for many other science cases as well. 

The quality of the light curves is not affected by changing the distribution of filters weights among the visits and remains roughly constant for all \texttt{filterdist} observing strategies.

\subsubsection{Rolling cadence}

Rolling observing strategies limit the visits to smaller regions of the sky. The Alt Scheduler algorithm (\texttt{alt} strategies in Figure \ref{fig:metricvalues}) chooses a north or south region of the sky to observe each night, which results in a larger portion of visits to a field being separated by $2$ nights (while longer time gaps are reduced). The rolling cadences enhance the number of visits to a chosen area on the sky (typically the sky is divided into several declination bands) in a given year, which results in a shorter revisit time to a field in the same year, at the expense of a longer revisit time in other years. This results in fewer transients being detected, because they are missed, when the field is not ``active'', but during the active periods, the light curves are more densely sampled. 

In the bottom panel of Figure \ref{fig:metricvalues} we show the results for the rolling cadences. As mentioned, the \texttt{alt} strategies add a nightly modulation to observations, varying between the Northern and Southern portions of the sky. The \texttt{alt\_roll} adds a $2$-strip rolling cadence. The \texttt{rolling\_mod2} strategy is a ``normal'' rolling cadence without the N-S nightly modulation and it devotes $6$ years to obtaining higher number of observations in two strips on the sky. The rest of the rolling observing strategies in the bottom panel of Figure \ref{fig:metricvalues} devote the first and the last $1.5$ years of the survey time to uniform coverage of the sky (as in the baseline), and in the remaining time the sky is split into $3$ declination bands. These strips are covered every third year. In some simulations, the strips are further split into $2$ bands in North and South, and an every-other nightly modulation between the Northern and Southern band is added. Simulations starting with \texttt{rolling\_nm} (nm = no modulation) do not add this modulation. The weight of the rolling cadence varies from $20$\% (scale $0.2$) to $100$\% (scale $1.0$), where a larger weight results in more visits in the emphasized declination band and fewer visits outside this band.

Our results show that the effect of rolling or varying the observations between the Northern/Southern portions of the sky is not reflected in the fraction of detected TDEs with \texttt{some\_color} and \texttt{some\_color\_pu} metrics, since they do not scale with the number of data points in the light curve. We see a trend toward worse performance, if the rolling is included, since this reduces the number of detected transients. 

The light curve quality scores, however, certainly benefit from the rolling cadence. Except for \texttt{alt\_dust} strategy (which is actually not rolling, but just observes either a Southern or Northern region of the sky each night), the light curve quality score is either not changed much or is increased. This is a perfect example of the trade-off between the number of detected TDEs (or the fraction, which is reduced) and the light curve quality (which is higher), if the temporal sampling is more frequent. Rolling cadence should therefore be better for transient science; we get better sampling over a smaller area. If a $10$\% decrease in the number of detected TDEs results in a $10$\% increase in the light curve score, we can treat this as a desirable trade. 

As mentioned in the end of \ref{subsec:intranight}, the increase in light curve quality score does not always imply a better quality of data, as is in the case of \texttt{baseline\_samefilt} observing strategy. When the light curve quality score increases, we can differentiate between two different scenarios: i) the light curve quality score increased because TDEs became better sampled, and ii) the light curve quality score increased because most of the TDEs that were missed are the ones that had the lowest light curve quality scores before and, on average, the remaining TDEs have a higher light curve quality score. In case of rolling cadences, the number of light curve points increases for a given TDE, which represents a genuine improvement in data quality rather than the fake improvement seen in \texttt{baseline\_samefilt}.

\section{Conclusions}
\label{sec:Conclusion}

The potential of Rubin Observatory for TDE observations is very promising, however, we expect that identifying TDEs out of a large number of transients will be challenging. In this work we used a sample of TDE light curves and simulated their observations with different observing strategies of the LSST using the MAF \texttt{TDEspop} metrics family. The metrics return as an output the recovered light curves and four different metric values: \texttt{some\_color}, which measures how effectively the color evolution of a TDE can be followed, \texttt{some\_color\_pu}, which is in addition sensitive to observations in $u$-band, which are important for photometric identification of TDEs, or \texttt{some\_color\_quality} and \texttt{some\_color\_\_pu\_quality}, which measure how well are the light curves, which passed the metric requirements of the \texttt{some\_color} and \texttt{some\_color\_pu} metrics, sampled in time. With these output metric values we investigate how varying a certain survey strategy parameter of the Rubin's observing strategy impacts the observations of TDEs. Our main findings are:
\begin{itemize}
    \item From the recovered light curves and light curve quality scores, we find that events which passed our color metrics will be generally sparsely sampled. In the \texttt{baseline} observing strategies, we can expect, on average, one data point every $4-5$ days. The light curve quality score does not vary significantly between different observing strategies and is at most improved by $10$\%.
    \item Pairs of visits to the same field on the sky in a single night are not strictly necessary for identification of TDEs, since colors (important for classification) can also be measured from fitting models to non-simultaneous flux measurements in different filters. Since the visit pairs are important for Solar System science and fast transients, we realize they will most likely be adopted in the survey and stress that in that case, pairs should be taken in different filters, otherwise it is impossible to measure colors near the peak of the light curve. Having a reliable color measurement near the peak of the light curve is important for classification of all extra-galactic transients.
    \item A smaller footprint of the survey (or a reduction in time allocated to mini-surveys not focused on transients) would provide denser temporal sampling, which is important for observing characteristics in light curves and for determining physical parameters of TDEs through model fitting.
    \item The increase of $u$-band exposure time is beneficial to reduce the effect of the readout noise, however, we stress that in that case, the longer $u$-band exposures should not come at a cost of lowering the number of $u$-band visits per field. As we have shown, this would be disastrous for TDE photometric identification relying on $u$-band observations.
    \item It is certainly beneficial for TDEs and most other extra-galactic transients if the filter weight distribution is skewed more towards bluer bands, since strategies with uniform filter distribution among visits or strategies with heavier weights on bluer filters will be significantly better for photometric identification of TDEs. Furthermore, including more observations in $u$ and $g$ band does not negatively impact other science cases.
    \item Rolling cadences provide better temporal sampling of TDEs and at the same time only decrease the number of detected TDEs by $\approx 10$\%. Having slightly fewer TDEs with good light curve coverage is certainly better than having a large number of poorly-sampled TDEs.
\end{itemize}

To conclude, we summarize what we consider as the best Rubin cadence for TDEs. An ideal survey would prioritize a rolling cadence to get denser sampling, with intra-night observations taken in different filters. A smaller footprint would be preferred to save observing time, and the increase in the observing time could be used to increase the number of observations in the $u$ and $g$ bands. The longer $u$-band exposures of the cadence should not come at the expense of a reduced $u$-band cadence.

In this work, we have not explicitly differentiated the contribution to the metric from the faint/fast TDE (iPTF-16fnl) or the ``normal" TDE lightcurve that we simulated (PS-10jh). For a flux-limited survey such as LSST, the fainter TDEs only contribute to the source counts at lower redshift; in our analysis we have no iPTF-16fnl detections above redshift $0.2$. A more detailed analysis on how certain survey strategies are biased towards different types of events at different redshifts would need to include a broader ensemble of different TDE types and a finer redshift grid. Such work is beyond the scope of this paper.

\begin{acknowledgments}
We thank Lynne Jones and Peter Yoachim for useful discussions and for their help in implementing the TDE metric into MAF. This paper was created in the Rubin LSST Transient and Variable Star (TVS) Science Collaboration (\url{https://lsst-tvssc.github.io/}). The authors acknowledge the support of Vera C. Rubin Legacy Survey of Space and Time Transient and Variable Stars Science Collaboration that provided opportunities for collaboration and exchange of ideas and knowledge of Rubin Observatory in the creation and implementation of this work. The authors also acknowledge the support of the LSST Corporation, which enabled the organization of workshops and hackathons throughout the cadence optimization process by directing private funding to these activities.
MN is supported by the European Research Council (ERC) under the European Union’s Horizon 2020 research and innovation programme (grant agreement No.~948381) and by a Fellowship from the Alan Turing Institute.
\end{acknowledgments}

\begin{acknowledgments}
KBB and AG acknowledge the financial support from the Slovenian Research Agency (research core funding P1-0031, infrastructure program I0-0033, project grant No. J1-8136, J1-2460 and KB’s Young Researcher grant) and networking support by the COST Action GWverse CA1610. AG acknowledges the support from the Fulbright Visiting Scholars program. 

\end{acknowledgments}

%

\facilities{Rubin Observatory}


\software{LSST metrics analysis framework \citep{Jones:2014}, \texttt{MOSFiT} \citep{Guillochon:2017bmg}, python packages numpy \citep{harris2020array} and plotly \citep{plotly}}



\appendix

\section{Appendix: Variations of observing strategies}
\label{appendix:a}

A starting point of all survey strategy simulations is the baseline observing strategy, which serves as a reference strategy. The baseline strategy is driven by the basic LSST science goals (see \citealt{Ivezic:2019ApJ} and \citealt{Bianco:2022ApJS}) and it consists of the main survey (wide-fast-deep; WFD) and four mini-surveys: Deep Drilling Field mini surveys (DDFs), the Galactic Plane mini survey (GP), the North Ecliptic Spur mini survey and the South Celestial Pole mini survey \citep{Ivezic:2019ApJ}. The most important survey for TDEs is the main survey, hence we mostly explore variations to the WFD in this work.

The WFD typically covers $\sim 18,000$ deg$^2$ of the sky between declinations of $-62^\circ <$ Dec $<+2^\circ$, excluding the central portion of the Galactic plane. Each night, two visits are acquired per field (of size $9.6$ deg$^2$), where a visit is defined either as two exposures of $15$ s each or one exposure of $30$ s, depending on the FBS version of the baseline strategy. The pair of visits is typically obtained in two different filters (the combinations are typically in adjacent filters: $u$-$g$, $g$-$r$, $r$-$i$, $i$-$z$, $z$-$y$, and $y$-$y$), and the visits are separated by $22$ min. This allows for identification of moving objects, fast transients, and improves the performance of the alert system. Visit pairs are then repeated every three to four nights (when the field is visible) and each field receives about $800$ visits in $10$ years, summed in all six filters. The fractions of observing time allocated per band are $0.06$, $0.09$, $0.22$, $0.22$, $0.20$, $0.21$ in $u$, $g$, $r$, $i$, $z$ and $y$, respectively \citep{lsstSRD}.

The baseline observing strategy serves as a starting point in the choice for the observing plan of Rubin. Various modifications to the baseline can lead to an increase of the scientific output of the survey, while the modified survey at the same time still obeys the defined scientific requirements it needs to meet. In Table \ref{table:strategies} we shortly describe what modifications were applied to each of the observing strategies we show in Figure \ref{fig:metricvalues}. More information about all available observing strategies can be found in \cite{Jones2020}, at the Vera C. Rubin Observatory LSST Community pages \footnote{\url{https://community.lsst.org/}} and at LSST Project Science Team's github pages\footnote{\url{https://github.com/lsst-pst}}.

\begin{center}
\begin{longtable}{p{.3\textwidth} p{.1\textwidth} p{.5\textwidth} }
\caption{Descriptions of survey strategy variations shown in Figure \ref{fig:metricvalues}.} \label{table:strategies} \\

\hline \multicolumn{1}{l}{Survey strategy} & \multicolumn{1}{l}{FBS version} & \multicolumn{1}{l}{Description} \\ \hline 
\endfirsthead

\multicolumn{3}{c}%
{{\bfseries \tablename\ \thetable{} -- continued from previous page}} \\
\hline \multicolumn{1}{l}{Survey strategy} & \multicolumn{1}{l}{FBS version} & \multicolumn{1}{l}{Description} \\ \hline 
\endhead

\hline
\endlastfoot

\texttt{baseline}                   & 1.5               & A visit to a given field in the sky is defined as $1 \times 30$ s exposure, which is a default visit for all FBS 1.5 strategies unless stated otherwise. \\

\texttt{baseline\_2snaps}           & 1.5               & Instead of $1 \times 30$ s exposure per visit, the visit consists of $2 \times 15$ s exposures. \\

\texttt{baseline\_samefilt}         & 1.5               & The two visits per field in a given night are obtained in the same filter. \\

\texttt{var\_expt}                  & 1.5               & The exposure times in a visit is varied based on the observing conditions. In good conditions, a visit is defined as $1 \times 20$ s exposure, while in poor conditions it can be extended up to $1\times 100$ s. The mean visit time is $32.2$ s, which results in $6$\% fewer overall number of visits when compared to the baseline.\\

\texttt{third\_obs\_pt15}           & 1.5               & A third visit to a given field in WFD is added at the end of the night, in either $g$, $r$, $i$ or $z$ filters. The time dedicated for obtaining the third visit is $15$ min per night.\\

\texttt{third\_obs\_pt30}           & 1.5               & A third visit to a given field in WFD is added at the end of the night, in either $g$, $r$, $i$ or $z$ filters. The time dedicated for obtaining the third visit is $30$ min per night.\\

\texttt{third\_obs\_pt60}           & 1.5               & A third visit to a given field in WFD is added at the end of the night, in either $g$, $r$, $i$ or $z$ filters. The time dedicated for obtaining the third visit is $60$ min per night.\\

\texttt{third\_obs\_pt90}           & 1.5               & A third visit to a given field in WFD is added at the end of the night, in either $g$, $r$, $i$ or $z$ filters. The time dedicated for obtaining the third visit is $90$ min per night.\\

\texttt{wfd\_scale0.70\_noddf}      & 1.5               & The fraction of time dedicated to the WFD area is $70$\% (i.e. the number of visits per field in the WFD is changed). DDF mini surveys are excluded, therefore the time allocated for WFD observations is increased by $5$\% when compared to the baseline.\\

\texttt{wfd\_scale0.80\_noddf}      & 1.5               & The fraction of time dedicated to the WFD area is $80$\%. DDF mini surveys are excluded, therefore the time allocated for WFD observations is increased by $5$\% when compared to the baseline.\\

\texttt{wfd\_scale0.90\_noddf}      & 1.5               & The fraction of time dedicated to the WFD area is $90$\%. DDF mini surveys are excluded, therefore the time allocated for WFD observations is increased by $5$\% when compared to the baseline.\\

\texttt{wfd\_scale0.99\_noddf}      & 1.5               & The fraction of time dedicated to the WFD area is $99$\%. DDF mini surveys are excluded, therefore the time allocated for WFD observations is increased by $5$\% when compared to the baseline.\\

\texttt{wfd\_scale0.70}             & 1.5               & The fraction of time dedicated to the WFD area is $70$\%. DDF mini surveys are included.\\

\texttt{wfd\_scale0.80}             & 1.5               & The fraction of time dedicated to the WFD area is $80$\%. DDF mini surveys are included.\\

\texttt{wfd\_scale0.90}             & 1.5               & The fraction of time dedicated to the WFD area is $90$\%. DDF mini surveys are included.\\

\texttt{wfd\_scale0.99}             & 1.5               & The fraction of time dedicated to the WFD area is $99$\%. DDF mini surveys are included.\\

\texttt{u60}                        & 1.5               & The default visit in $u$-band consists of $1\times60$ s exposure, which results in reducing the total number of $u$-band visits to a given field in the sky by $50$\%.\\

Uniform (\texttt{filterdist\_indx1})    & 1.5           & The distribution of observing time allocated per band is uniform.\\

Baseline-like (\texttt{filterdist\_indx2})  & 1.5       & The distribution of observing time allocated per band is similar to the baseline strategy.\\

$g$ heavy (\texttt{filterdist\_indx3})      & 1.5       & Additional $10$\% of observing time is allocated for observations in $g$-band at the expense of reduced time allocated for observations in $r$, $i$, $z$ and $y$ bands when compared to the baseline. \\

$u$ heavy (\texttt{filterdist\_indx4})      & 1.5       & Additional $10$\% of observing time is allocated for observations in $u$-band at the expense of reduced time allocated for observations in $r$, $i$, $z$ and $y$ bands when compared to the baseline.\\

$z$ and $y$ heavy (\texttt{filterdist\_indx5}) & 1.5     & Additional $6$\% of observing time is allocated for observations in $z$ and $y$ bands, respectively, at the expense of reduced time allocated for observations in $u$, $g$, $r$ and $i$ bands when compared to the baseline.\\

$i$ heavy (\texttt{filterdist\_indx6})         & 1.5     & Additional $7$\% of observing time is allocated for observations in $i$ band at the expense of reduced time allocated for observations in $g$, $r$, $z$ and $y$ bands when compared to the baseline.\\

Bluer (\texttt{filterdist\_indx7})             & 1.5     & Additional $3$\% of observing time is allocated for observations in $u$ and $g$ bands, respectively, at the expense of reduced time allocated for observations in $r$, $i$, $z$ and $y$ bands when compared to the baseline.\\

Redder (\texttt{filterdist\_indx8})            & 1.5     & Additional $1$\% of observing time is allocated for observatoins in $z$ band and additional $2$\% of observing time is allocated for observations in $y$ band, keeping the time allocated for observations in $i$ band the same as in the baseline. When compared to the baseline, the time allocated for observations in $g$ and $r$ bands is decreased.\\

\texttt{alt\_dust}                  & 1.5                & Visits alternate between Northern and Southern portions of the sky in the WFD on a nightly basis. This adds a night-long gap between revisits to a field.\\

\texttt{alt\_roll\_mod2\_dust\_sdf\_0.20}      & 1.5     & The survey is executed in non-uniform manner, where some regions of the sky receive a higher number of visits over a defined season, followed by a lower number of visits in the next season. The sky in this strategy is split into $2$ declination bands and a total of 6 years are devoted for rolling (i. e. for obtaining a higher number of observations in the two declination bands). A every-other nightly modulation between the northern and the southern sub-section of each declination band is added.\\

\texttt{roll\_mod2\_dust\_sdf\_0.20}           & 1.5    & The survey is executed in non-uniform manner, where some regions of the sky receive a higher number of visits over a defined season, followed by a lower number of visits in the next season. The sky in this strategy is split into $2$ declination bands and a total of 6 years are devoted for rolling (i. e. for obtaining a higher number of observations in the two declination bands).\\

\texttt{baseline\_nexp1}            & 1.7               & Instead of $2 \times 15$ s exposures per visit, the visit consists of $1 \times 30$ s exposure.\\

\texttt{baseline\_nexp2}            & 1.7               & A visit to a given field in the sky is defined as $2 \times 15$ s exposures, which is a default visit for all FBS 1.7 strategies unless stated otherwise.\\

\texttt{pair\_times\_11}            & 1.7               & The time between pairs of visits in a night is $11$ min instead of $22$ min. \\

\texttt{pair\_times\_44}            & 1.7               & The time between pairs of visits in a night is $44$ min instead of $22$ min. \\

\texttt{pair\_times\_55}            & 1.7               & The time between pairs of visits in a night is $55$ min instead of $22$ min.\\

\texttt{footprint\_1}               & 1.7               & The traditional WFD footprint is increased to $-70.2 < $ Dec $< 7.8$ (by approx $2000$ deg$^2$) in order to cover the galactic bulge and Magellanic clouds. The Northern and Southern limits are set by dust-extinction, while the coverage on the remaining sky is varied.\\

\texttt{footprint\_2}               & 1.7               & The traditional WFD footprint is increased to $-67.4 < $ Dec $< 8$ (by approx $2000$ deg$^2$) in order to cover the galactic bulge and Magellanic clouds. The Northern and Southern limits are set by dust-extinction, while the coverage on the remaining sky is varied. \\

\texttt{footprint\_3}               & 1.7               & The traditional WFD footprint is increased to $-67.4 < $ Dec $< 8$ (by approx $2000$ deg$^2$) in order to cover the galactic bulge and Magellanic clouds. The Northern and Southern limits are set by dust-extinction, while the coverage on the remaining sky is varied. An additional $20$ deg band in declination is added to WFD to cover a bridge across the Galactic Plane at higher galactic latitudes. \\

\texttt{footprint\_5}               & 1.7               & The traditional WFD footprint is increased to $-67.4 < $ Dec $< 8$ (by approx $2000$ deg$^2$) in order to cover the galactic bulge and Magellanic clouds. The Northern and Southern limits are set by dust-extinction, while the coverage on the remaining sky is varied. An additional $20$ deg band in declination is added to WFD to cover a bridge across the Galactic Plane at higher galactic latitudes.\\

\texttt{footprint\_6}               & 1.7               & The traditional WFD footprint is increased to $-67.4 < $ Dec $< 8$ (by approx $2000$ deg$^2$) in order to cover the galactic bulge and Magellanic clouds. The Northern and Southern limits are set by dust-extinction, while the coverage on the remaining sky is varied. An additional $20$ deg band in declination is added to WFD to cover a bridge across the Galactic Plane at higher galactic latitudes.\\

\texttt{u\_long\_ms\_30}            & 1.7               & The default visit in $u$-band consists of $1\times30$ s exposure, while the number of $u$ band visits is left unchanged. This results in reducing the total number of visits in the remaining bands.\\

\texttt{u\_long\_ms\_40}            & 1.7               & The default visit in $u$-band consists of $1\times40$ s exposure, while the number of $u$ band visits is left unchanged. This results in reducing the total number of visits in the remaining bands.\\

\texttt{u\_long\_ms\_50}            & 1.7               & The default visit in $u$-band consists of $1\times50$ s exposure, while the number of $u$ band visits is left unchanged. This results in reducing the total number of visits in the remaining bands.\\

\texttt{u\_long\_ms\_60}            & 1.7               & The default visit in $u$-band consists of $1\times60$ s exposure, while the number of $u$ band visits is left unchanged. This results in reducing the total number of visits in the remaining bands. \\

\texttt{rolling\_nm\_scale0.2\_nslice3}        & 1.7    & The survey is executed in non-uniform manner, where some regions of the sky receive a higher number of visits over a defined season, followed by a lower number of visits in the next season. The sky in this strategy is split into $3$ and a total of $1.2$ years are devoted for rolling (i. e. for obtaining a higher number of observations in the two declination bands). A every-other nightly modulation between the northern and the southern sub-section of each declination band is added.\\

\texttt{rolling\_nm\_scale0.4\_nslice3}        & 1.7     & The survey is executed in non-uniform manner, where some regions of the sky receive a higher number of visits over a defined season, followed by a lower number of visits in the next season. The sky in this strategy is split into $3$ and a total of $2.4$ years are devoted for rolling (i. e. for obtaining a higher number of observations in the two declination bands). A every-other nightly modulation between the northern and the southern sub-section of each declination band is added.\\

\texttt{rolling\_nm\_scale0.6\_nslice3}        & 1.7      & The survey is executed in non-uniform manner, where some regions of the sky receive a higher number of visits over a defined season, followed by a lower number of visits in the next season. The sky in this strategy is split into $3$ and a total of $3.6$ years are devoted for rolling (i. e. for obtaining a higher number of observations in the two declination bands). A every-other nightly modulation between the northern and the southern sub-section of each declination band is added.\\

\texttt{rolling\_nm\_scale0.8\_nslice3}        & 1.7      & The survey is executed in non-uniform manner, where some regions of the sky receive a higher number of visits over a defined season, followed by a lower number of visits in the next season. The sky in this strategy is split into $3$ and a total of $4.8$ years are devoted for rolling (i. e. for obtaining a higher number of observations in the two declination bands). A every-other nightly modulation between the northern and the southern sub-section of each declination band is added.\\

\texttt{rolling\_nm\_scale0.9\_nslice3}        & 1.7       & The survey is executed in non-uniform manner, where some regions of the sky receive a higher number of visits over a defined season, followed by a lower number of visits in the next season. The sky in this strategy is split into $3$ and a total of $5.4$ years are devoted for rolling (i. e. for obtaining a higher number of observations in the two declination bands). A every-other nightly modulation between the northern and the southern sub-section of each declination band is added.\\

\texttt{rolling\_nm\_scale1.0\_nslice3}        & 1.7      & The survey is executed in non-uniform manner, where some regions of the sky receive a higher number of visits over a defined season, followed by a lower number of visits in the next season. The sky in this strategy is split into $3$ and a total of $6$ years are devoted for rolling (i. e. for obtaining a higher number of observations in the two declination bands). A every-other nightly modulation between the northern and the southern sub-section of each declination band is added.\\

\texttt{rolling\_scale0.2\_nslice3}            & 1.7      & The survey is executed in non-uniform manner, where some regions of the sky receive a higher number of visits over a defined season, followed by a lower number of visits in the next season. The sky in this strategy is split into $3$ and a total of $1.2$ years are devoted for rolling (i. e. for obtaining a higher number of observations in the two declination bands).\\

\texttt{rolling\_scale0.4\_nslice3}            & 1.7        & The survey is executed in non-uniform manner, where some regions of the sky receive a higher number of visits over a defined season, followed by a lower number of visits in the next season. The sky in this strategy is split into $3$ and a total of $2.4$ years are devoted for rolling (i. e. for obtaining a higher number of observations in the two declination bands).\\

\texttt{rolling\_scale0.6\_nslice3}            & 1.7      & The survey is executed in non-uniform manner, where some regions of the sky receive a higher number of visits over a defined season, followed by a lower number of visits in the next season. The sky in this strategy is split into $3$ and a total of $3.6$ years are devoted for rolling (i. e. for obtaining a higher number of observations in the two declination bands).\\

\texttt{rolling\_scale0.8\_nslice3}            & 1.7     & The survey is executed in non-uniform manner, where some regions of the sky receive a higher number of visits over a defined season, followed by a lower number of visits in the next season. The sky in this strategy is split into $3$ and a total of $4.8$ years are devoted for rolling (i. e. for obtaining a higher number of observations in the two declination bands).\\

\texttt{rolling\_scale0.9\_nslice3}            & 1.7      & The survey is executed in non-uniform manner, where some regions of the sky receive a higher number of visits over a defined season, followed by a lower number of visits in the next season. The sky in this strategy is split into $3$ and a total of $5.4$ years are devoted for rolling (i. e. for obtaining a higher number of observations in the two declination bands).\\

\texttt{rolling\_scale1.0\_nslice3}            & 1.7       & The survey is executed in non-uniform manner, where some regions of the sky receive a higher number of visits over a defined season, followed by a lower number of visits in the next season. The sky in this strategy is split into $3$ and a total of $6$ years are devoted for rolling (i. e. for obtaining a higher number of observations in the two declination bands).\\

\hline

\end{longtable}
\end{center}


\bibliography{references}{}

\begin{thebibliography}{}
\expandafter\ifx\csname natexlab\endcsname\relax\def\natexlab#1{#1}\fi
\providecommand{\url}[1]{\href{#1}{#1}}
\providecommand{\dodoi}[1]{doi:~\href{http://doi.org/#1}{\nolinkurl{#1}}}
\providecommand{\doeprint}[1]{\href{http://ascl.net/#1}{\nolinkurl{http://ascl.net/#1}}}
\providecommand{\doarXiv}[1]{\href{https://arxiv.org/abs/#1}{\nolinkurl{https://arxiv.org/abs/#1}}}

\bibitem[{{Bellm} {et~al.}(2019){Bellm}, {Kulkarni}, {Barlow}, {Feindt},
  {Graham}, {Goobar}, {Kupfer}, {Ngeow}, {Nugent}, {Ofek}, {Prince}, {Riddle},
  {Walters}, \& {Ye}}]{Bellm:2019PASP}
{Bellm}, E.~C., {Kulkarni}, S.~R., {Barlow}, T., {et~al.} 2019, PASP, 131,
  068003, \dodoi{10.1088/1538-3873/ab0c2a}

\bibitem[{{Bianco} {et~al.}(2022){Bianco}, {Ivezi{\'c}}, {Jones}, {Graham},
  {Marshall}, {Saha}, {Strauss}, {Yoachim}, {Ribeiro}, {Anguita}, {Bauer},
  {Bauer}, {Bellm}, {Blum}, {Brandt}, {Brough}, {Catelan}, {Clarkson},
  {Connolly}, {Gawiser}, {Gizis}, {Hlo{\v{z}}ek}, {Kaviraj}, {Liu}, {Lochner},
  {Mahabal}, {Mandelbaum}, {McGehee}, {Neilsen}, {Olsen}, {Peiris}, {Rhodes},
  {Richards}, {Ridgway}, {Schwamb}, {Scolnic}, {Shemmer}, {Slater}, {Slosar},
  {Smartt}, {Strader}, {Street}, {Trilling}, {Verma}, {Vivas}, {Wechsler}, \&
  {Willman}}]{Bianco:2022ApJS}
{Bianco}, F.~B., {Ivezi{\'c}}, {\v{Z}}., {Jones}, R.~L., {et~al.} 2022, \apjs,
  258, 1, \dodoi{10.3847/1538-4365/ac3e72}

\bibitem[{{Blagorodnova} {et~al.}(2017){Blagorodnova}, {Gezari}, {Hung},
  {Kulkarni}, {Cenko}, {Pasham}, {Yan}, {Arcavi}, {Ben-Ami}, {Bue}, {Cantwell},
  {Cao}, {Castro-Tirado}, {Fender}, {Fremling}, {Gal-Yam}, {Ho}, {Horesh},
  {Hosseinzadeh}, {Kasliwal}, {Kong}, {Laher}, {Leloudas}, {Lunnan}, {Masci},
  {Mooley}, {Neill}, {Nugent}, {Powell}, {Valeev}, {Vreeswijk}, {Walters}, \&
  {Wozniak}}]{Blagorodnova:2017ApJ}
{Blagorodnova}, N., {Gezari}, S., {Hung}, T., {et~al.} 2017, ApJ, 844, 46,
  \dodoi{10.3847/1538-4357/aa7579}

\bibitem[{{Bonnerot} {et~al.}(2017){Bonnerot}, {Rossi}, \&
  {Lodato}}]{Bonnerot:2017MNRAS}
{Bonnerot}, C., {Rossi}, E.~M., \& {Lodato}, G. 2017, MNRAS, 464, 2816,
  \dodoi{10.1093/mnras/stw2547}

\bibitem[{{Bricman} \& {Gomboc}(2020)}]{Bricman:2020ApJ}
{Bricman}, K., \& {Gomboc}, A. 2020, ApJ, 890, 73,
  \dodoi{10.3847/1538-4357/ab6989}

\bibitem[{{Delgado} {et~al.}(2014){Delgado}, {Saha}, {Chandrasekharan}, {Cook},
  {Petry}, \& {Ridgway}}]{Delgado:2014SPIE}
{Delgado}, F., {Saha}, A., {Chandrasekharan}, S., {et~al.} 2014, in {Society of
  Photo-Optical Instrumentation Engineers (SPIE) Conference Series}, Vol.
  {9150}, Modeling, Systems Engineering, and Project Management for Astronomy
  VI, ed. G.~Z. {Angeli} \& P.~{Dierickx}, {915015},
  \dodoi{{10.1117/12.2056898}}

\bibitem[{{Drout} {et~al.}(2014){Drout}, {Chornock}, {Soderberg}, {Sanders},
  {McKinnon}, {Rest}, {Foley}, {Milisavljevic}, {Margutti}, {Berger},
  {Calkins}, {Fong}, {Gezari}, {Huber}, {Kankare}, {Kirshner}, {Leibler},
  {Lunnan}, {Mattila}, {Marion}, {Narayan}, {Riess}, {Roth}, {Scolnic},
  {Smartt}, {Tonry}, {Burgett}, {Chambers}, {Hodapp}, {Jedicke}, {Kaiser},
  {Magnier}, {Metcalfe}, {Morgan}, {Price}, \& {Waters}}]{Drout:2014ApJ}
{Drout}, M.~R., {Chornock}, R., {Soderberg}, A.~M., {et~al.} 2014, ApJ, 794,
  23, \dodoi{10.1088/0004-637X/794/1/23}

\bibitem[{Evans \& Kochanek(1989)}]{Evans:1989qe}
Evans, C.~R., \& Kochanek, C.~S. 1989, ApJ, 346, L13, \dodoi{10.1086/185567}

\bibitem[{{Gezari}(2021)}]{Gezari2021ARAA}
{Gezari}, S. 2021, \araa, 59, 21, \dodoi{10.1146/annurev-astro-111720-030029}

\bibitem[{{Gezari} {et~al.}(2012){Gezari}, {Chornock}, {Rest}, {Huber},
  {Forster}, {Berger}, {Challis}, {Neill}, {Martin}, {Heckman}, {Lawrence},
  {Norman}, {Narayan}, {Foley}, {Marion}, {Scolnic}, {Chomiuk}, {Soderberg},
  {Smith}, {Kirshner}, {Riess}, {Smartt}, {Stubbs}, {Tonry}, {Wood-Vasey},
  {Burgett}, {Chambers}, {Grav}, {Heasley}, {Kaiser}, {Kudritzki}, {Magnier},
  {Morgan}, \& {Price}}]{Gezari:2012Nat}
{Gezari}, S., {Chornock}, R., {Rest}, A., {et~al.} 2012, Nature, 485, 217,
  \dodoi{10.1038/nature10990}

\bibitem[{Gomboc \& {\v C}ade{\v z}(2005)}]{Gomboc:2005wu}
Gomboc, A., \& {\v C}ade{\v z}, A. 2005, ApJ, 625, 278, \dodoi{10.1086/429263}

\bibitem[{Guillochon {et~al.}(2018)Guillochon, Nicholl, Villar, Mockler,
  Narayan, Mandel, Berger, \& Williams}]{Guillochon:2017bmg}
Guillochon, J., Nicholl, M., Villar, V.~A., {et~al.} 2018, ApJS, 236, 6,
  \dodoi{10.3847/1538-4365/aab761}

\bibitem[{Guillochon \& Ramirez-Ruiz(2013)}]{Guillochon:2012uc}
Guillochon, J., \& Ramirez-Ruiz, E. 2013, ApJ, 767, 25,
  \dodoi{10.1088/0004-637X/798/1/64, 10.1088/0004-637X/767/1/25}

\bibitem[{{Guillochon} \& {Ramirez-Ruiz}(2013)}]{Guillochon:2013ApJ}
{Guillochon}, J., \& {Ramirez-Ruiz}, E. 2013, ApJ, 767, 25,
  \dodoi{10.1088/0004-637X/767/1/25}

\bibitem[{Harris {et~al.}(2020)Harris, Millman, van~der Walt, Gommers,
  Virtanen, Cournapeau, Wieser, Taylor, Berg, Smith, Kern, Picus, Hoyer, van
  Kerkwijk, Brett, Haldane, del R{\'{i}}o, Wiebe, Peterson,
  G{\'{e}}rard-Marchant, Sheppard, Reddy, Weckesser, Abbasi, Gohlke, \&
  Oliphant}]{harris2020array}
Harris, C.~R., Millman, K.~J., van~der Walt, S.~J., {et~al.} 2020, Nature, 585,
  357, \dodoi{10.1038/s41586-020-2649-2}

\bibitem[{{Hinkle} {et~al.}(2020){Hinkle}, {Holoien}, {Shappee}, {Auchettl},
  {Kochanek}, {Stanek}, {Payne}, \& {Thompson}}]{Hinkle2020ApJ}
{Hinkle}, J.~T., {Holoien}, T. W.~S., {Shappee}, B.~J., {et~al.} 2020, \apjl,
  894, L10, \dodoi{10.3847/2041-8213/ab89a2}

\bibitem[{{Holoien} {et~al.}(2016){Holoien}, {Kochanek}, {Prieto}, {Stanek},
  {Dong}, {Shappee}, {Grupe}, {Brown}, {Basu}, {Beacom}, {Bersier},
  {Brimacombe}, {Danilet}, {Falco}, {Guo}, {Jose}, {Herczeg}, {Long},
  {Pojmanski}, {Simonian}, {Szczygie{\l}}, {Thompson}, {Thorstensen}, {Wagner},
  \& {Wo{\'z}niak}}]{Holoien:2015pza}
{Holoien}, T.~W.~S., {Kochanek}, C.~S., {Prieto}, J.~L., {et~al.} 2016, MNRAS,
  455, 2918, \dodoi{10.1093/mnras/stv2486}

\bibitem[{{Hung} {et~al.}(2017){Hung}, {Gezari}, {Blagorodnova}, {Roth},
  {Cenko}, {Kulkarni}, {Horesh}, {Arcavi}, {McCully}, {Yan}, {Lunnan},
  {Fremling}, {Cao}, {Nugent}, \& {Wozniak}}]{Hung:2017ApJ}
{Hung}, T., {Gezari}, S., {Blagorodnova}, N., {et~al.} 2017, ApJ, 842, 29,
  \dodoi{10.3847/1538-4357/aa7337}

\bibitem[{{Ivezi{\'c}} {et~al.}(2019){Ivezi{\'c}}, {Kahn}, {Tyson}, {Abel},
  {Acosta}, {Allsman}, {Alonso}, {AlSayyad}, {Anderson}, {Andrew}, {Angel},
  {Angeli}, {Ansari}, {Antilogus}, {Araujo}, {Armstrong}, {Arndt}, {Astier},
  {Aubourg}, {Auza}, {Axelrod}, {Bard}, {Barr}, {Barrau}, {Bartlett}, {Bauer},
  {Bauman}, {Baumont}, {Bechtol}, {Bechtol}, {Becker}, {Becla}, {Beldica},
  {Bellavia}, {Bianco}, {Biswas}, {Blanc}, {Blazek}, {Bland ford}, {Bloom},
  {Bogart}, {Bond}, {Booth}, {Borgland}, {Borne}, {Bosch}, {Boutigny},
  {Brackett}, {Bradshaw}, {Brand t}, {Brown}, {Bullock}, {Burchat}, {Burke},
  {Cagnoli}, {Calabrese}, {Callahan}, {Callen}, {Carlin}, {Carlson}, {Chand
  rasekharan}, {Charles-Emerson}, {Chesley}, {Cheu}, {Chiang}, {Chiang},
  {Chirino}, {Chow}, {Ciardi}, {Claver}, {Cohen-Tanugi}, {Cockrum}, {Coles},
  {Connolly}, {Cook}, {Cooray}, {Covey}, {Cribbs}, {Cui}, {Cutri}, {Daly},
  {Daniel}, {Daruich}, {Daubard}, {Daues}, {Dawson}, {Delgado}, {Dellapenna},
  {de Peyster}, {de Val-Borro}, {Digel}, {Doherty}, {Dubois},
  {Dubois-Felsmann}, {Durech}, {Economou}, {Eifler}, {Eracleous}, {Emmons},
  {Fausti Neto}, {Ferguson}, {Figueroa}, {Fisher-Levine}, {Focke}, {Foss},
  {Frank}, {Freemon}, {Gangler}, {Gawiser}, {Geary}, {Gee}, {Geha}, {Gessner},
  {Gibson}, {Gilmore}, {Glanzman}, {Glick}, {Goldina}, {Goldstein}, {Goodenow},
  {Graham}, {Gressler}, {Gris}, {Guy}, {Guyonnet}, {Haller}, {Harris},
  {Hascall}, {Haupt}, {Hernand ez}, {Herrmann}, {Hileman}, {Hoblitt},
  {Hodgson}, {Hogan}, {Howard}, {Huang}, {Huffer}, {Ingraham}, {Innes},
  {Jacoby}, {Jain}, {Jammes}, {Jee}, {Jenness}, {Jernigan}, {Jevremovi{\'c}},
  {Johns}, {Johnson}, {Johnson}, {Jones}, {Juramy-Gilles}, {Juri{\'c}},
  {Kalirai}, {Kallivayalil}, {Kalmbach}, {Kantor}, {Karst}, {Kasliwal},
  {Kelly}, {Kessler}, {Kinnison}, {Kirkby}, {Knox}, {Kotov}, {Krabbendam},
  {Krughoff}, {Kub{\'a}nek}, {Kuczewski}, {Kulkarni}, {Ku}, {Kurita}, {Lage},
  {Lambert}, {Lange}, {Langton}, {Le Guillou}, {Levine}, {Liang}, {Lim},
  {Lintott}, {Long}, {Lopez}, {Lotz}, {Lupton}, {Lust}, {MacArthur}, {Mahabal},
  {Mand elbaum}, {Markiewicz}, {Marsh}, {Marshall}, {Marshall}, {May},
  {McKercher}, {McQueen}, {Meyers}, {Migliore}, {Miller}, {Mills}, {Miraval},
  {Moeyens}, {Moolekamp}, {Monet}, {Moniez}, {Monkewitz}, {Montgomery},
  {Morrison}, {Mueller}, {Muller}, {Mu{\~n}oz Arancibia}, {Neill}, {Newbry},
  {Nief}, {Nomerotski}, {Nordby}, {O'Connor}, {Oliver}, {Olivier}, {Olsen},
  {O'Mullane}, {Ortiz}, {Osier}, {Owen}, {Pain}, {Palecek}, {Parejko},
  {Parsons}, {Pease}, {Peterson}, {Peterson}, {Petravick}, {Libby Petrick},
  {Petry}, {Pierfederici}, {Pietrowicz}, {Pike}, {Pinto}, {Plante}, {Plate},
  {Plutchak}, {Price}, {Prouza}, {Radeka}, {Rajagopal}, {Rasmussen},
  {Regnault}, {Reil}, {Reiss}, {Reuter}, {Ridgway}, {Riot}, {Ritz}, {Robinson},
  {Roby}, {Roodman}, {Rosing}, {Roucelle}, {Rumore}, {Russo}, {Saha},
  {Sassolas}, {Schalk}, {Schellart}, {Schindler}, {Schmidt}, {Schneider},
  {Schneider}, {Schoening}, {Schumacher}, {Schwamb}, {Sebag}, {Selvy},
  {Sembroski}, {Seppala}, {Serio}, {Serrano}, {Shaw}, {Shipsey}, {Sick},
  {Silvestri}, {Slater}, {Smith}, {Smith}, {Sobhani}, {Soldahl},
  {Storrie-Lombardi}, {Stover}, {Strauss}, {Street}, {Stubbs}, {Sullivan},
  {Sweeney}, {Swinbank}, {Szalay}, {Takacs}, {Tether}, {Thaler}, {Thayer},
  {Thomas}, {Thornton}, {Thukral}, {Tice}, {Trilling}, {Turri}, {Van Berg},
  {Vanden Berk}, {Vetter}, {Virieux}, {Vucina}, {Wahl}, {Walkowicz}, {Walsh},
  {Walter}, {Wang}, {Wang}, {Warner}, {Wiecha}, {Willman}, {Winters},
  {Wittman}, {Wolff}, {Wood-Vasey}, {Wu}, {Xin}, {Yoachim}, \&
  {Zhan}}]{Ivezic:2019ApJ}
{Ivezi{\'c}}, {\v{Z}}., {Kahn}, S.~M., {Tyson}, J.~A., {et~al.} 2019, ApJ, 873,
  111, \dodoi{10.3847/1538-4357/ab042c}

\bibitem[{{Ivezi{\'c}, {\v{Z}.} and the LSST Science
  Collaboration}(2013)}]{lsstSRD}
{Ivezi{\'c}, {\v{Z}.} and the LSST Science Collaboration}. 2013, LSST Science
  Requirements Document.
\newblock \url{http://ls.st/LPM-17}

\bibitem[{{Jones} {et~al.}(2020){Jones}, {Yoachim}, {Ivezi\'{c}}, {Nielsen}, \&
  {Ribeiro}}]{Jones2020}
{Jones}, R.~L., {Yoachim}, P., {Ivezi\'{c}}, {\v{Z}}., {Nielsen}, E.~H., \&
  {Ribeiro}, T. 2020, {Survey Strategy and Cadence Choices for the Vera C.
  Rubin Observatory Legacy Survey of Space and Time (LSST) (v1.2). Zenodo.}
\newblock \url{https://doi.org/10.5281/zenodo.4048838}

\bibitem[{{Jones} {et~al.}(2014){Jones}, {Yoachim}, {Chandrasekharan},
  {Connolly}, {Cook}, {Ivezic}, {Krughoff}, {Petry}, \& {Ridgway}}]{Jones:2014}
{Jones}, R.~L., {Yoachim}, P., {Chandrasekharan}, S., {et~al.} 2014, in
  {Society of Photo-Optical Instrumentation Engineers (SPIE) Conference
  Series}, Vol. {9149}, {Observatory Operations: Strategies, Processes, and
  Systems V}, ed. {{Peck}, Alison B. and {Benn}, Chris R. and {Seaman}, Robert
  L.}, {91490B}, \dodoi{{10.1117/12.2056835}}

\bibitem[{Kochanek(1994)}]{Kochanek:1993cm}
Kochanek, C.~S. 1994, ApJ, 422, 508, \dodoi{10.1086/173745}

\bibitem[{{Lu} \& {Bonnerot}(2020)}]{Lu:2020MNRAS}
{Lu}, W., \& {Bonnerot}, C. 2020, MNRAS, 492, 686,
  \dodoi{10.1093/mnras/stz3405}

\bibitem[{{Margutti} {et~al.}(2019){Margutti}, {Metzger}, {Chornock}, {Vurm},
  {Roth}, {Grefenstette}, {Savchenko}, {Cartier}, {Steiner}, {Terreran},
  {Margalit}, {Migliori}, {Milisavljevic}, {Alexander}, {Bietenholz},
  {Blanchard}, {Bozzo}, {Brethauer}, {Chilingarian}, {Coppejans}, {Ducci},
  {Ferrigno}, {Fong}, {G{\"o}tz}, {Guidorzi}, {Hajela}, {Hurley}, {Kuulkers},
  {Laurent}, {Mereghetti}, {Nicholl}, {Patnaude}, {Ubertini}, {Banovetz},
  {Bartel}, {Berger}, {Coughlin}, {Eftekhari}, {Frederiks}, {Kozlova},
  {Laskar}, {Svinkin}, {Drout}, {MacFadyen}, \& {Paterson}}]{Margutti:2019ApJ}
{Margutti}, R., {Metzger}, B.~D., {Chornock}, R., {et~al.} 2019, ApJ, 872, 18,
  \dodoi{10.3847/1538-4357/aafa01}

\bibitem[{Mockler {et~al.}(2019)Mockler, Guillochon, \&
  Ramirez-Ruiz}]{Mockler:2018xne}
Mockler, B., Guillochon, J., \& Ramirez-Ruiz, E. 2019, ApJ., 872, 151,
  \dodoi{10.3847/1538-4357/ab010f}

\bibitem[{{Nicholl} {et~al.}(2022){Nicholl}, {Lanning}, {Ramsden}, {Mockler},
  {Lawrence}, {Short}, \& {Ridley}}]{Nicholl:2020arxiv}
{Nicholl}, M., {Lanning}, D., {Ramsden}, P., {et~al.} 2022, arXiv e-prints,
  arXiv:2201.02649.
\newblock \doarXiv{2201.02649}

\bibitem[{{Piran} {et~al.}(2015){Piran}, {Svirski}, {Krolik}, {Cheng}, \&
  {Shiokawa}}]{Piran:2015ApJ}
{Piran}, T., {Svirski}, G., {Krolik}, J., {Cheng}, R.~M., \& {Shiokawa}, H.
  2015, ApJ, 806, 164, \dodoi{10.1088/0004-637X/806/2/164}

\bibitem[{{Plotly Technologies Inc.}(2015)}]{plotly}
{Plotly Technologies Inc.} 2015, Collaborative data science,  Montreal, QC:
  Plotly Technologies Inc.
\newblock \url{https://plot.ly}

\bibitem[{Rees(1988)}]{Rees:1988bf}
Rees, M.~J. 1988, Nature, 333, 523, \dodoi{10.1038/333523a0}

\bibitem[{{Roth} {et~al.}(2016){Roth}, {Kasen}, {Guillochon}, \&
  {Ramirez-Ruiz}}]{Roth:2016ApJ}
{Roth}, N., {Kasen}, D., {Guillochon}, J., \& {Ramirez-Ruiz}, E. 2016, ApJ,
  827, 3, \dodoi{10.3847/0004-637X/827/1/3}

\bibitem[{{Shappee} {et~al.}(2014){Shappee}, {Prieto}, {Grupe}, {Kochanek},
  {Stanek}, {De Rosa}, {Mathur}, {Zu}, {Peterson}, {Pogge}, {Komossa}, {Im},
  {Jencson}, {Holoien}, {Basu}, {Beacom}, {Szczygie{\l}}, {Brimacombe},
  {Adams}, {Campillay}, {Choi}, {Contreras}, {Dietrich}, {Dubberley},
  {Elphick}, {Foale}, {Giustini}, {Gonzalez}, {Hawkins}, {Howell}, {Hsiao},
  {Koss}, {Leighly}, {Morrell}, {Mudd}, {Mullins}, {Nugent}, {Parrent},
  {Phillips}, {Pojmanski}, {Rosing}, {Ross}, {Sand}, {Terndrup}, {Valenti},
  {Walker}, \& {Yoon}}]{Shappee:2013mna}
{Shappee}, B.~J., {Prieto}, J.~L., {Grupe}, D., {et~al.} 2014, ApJ, 788, 48,
  \dodoi{10.1088/0004-637X/788/1/48}

\bibitem[{{Stone} {et~al.}(2020){Stone}, {Vasiliev}, {Kesden}, {Rossi},
  {Perets}, \& {Amaro-Seoane}}]{Stone2020SSRv}
{Stone}, N.~C., {Vasiliev}, E., {Kesden}, M., {et~al.} 2020, \ssr, 216, 35,
  \dodoi{10.1007/s11214-020-00651-4}

\bibitem[{{Tonry} {et~al.}(2018){Tonry}, {Denneau}, {Heinze}, {Stalder},
  {Smith}, {Smartt}, {Stubbs}, {Weiland }, \& {Rest}}]{Tonry:2018PASP}
{Tonry}, J.~L., {Denneau}, L., {Heinze}, A.~N., {et~al.} 2018, PASP, 130,
  064505, \dodoi{10.1088/1538-3873/aabadf}

\bibitem[{{van Velzen}(2018)}]{vanVelzen:2018ApJ}
{van Velzen}, S. 2018, \apj, 852, 72, \dodoi{10.3847/1538-4357/aa998e}

\bibitem[{{van Velzen} {et~al.}(2020){van Velzen}, {Holoien}, {Onori}, {Hung},
  \& {Arcavi}}]{vanVelzen:2020SSRv}
{van Velzen}, S., {Holoien}, T. W.~S., {Onori}, F., {Hung}, T., \& {Arcavi}, I.
  2020, SSR, 216, 124, \dodoi{10.1007/s11214-020-00753-z}

\bibitem[{{van Velzen} {et~al.}(2011){van Velzen}, {Farrar}, {Gezari},
  {Morrell}, {Zaritsky}, {{\"O}stman}, {Smith}, {Gelfand}, \&
  {Drake}}]{vanVelzen:2011ApJ}
{van Velzen}, S., {Farrar}, G.~R., {Gezari}, S., {et~al.} 2011, ApJ, 741, 73,
  \dodoi{10.1088/0004-637X/741/2/73}

\bibitem[{{van Velzen} {et~al.}(2019){van Velzen}, {Gezari}, {Cenko}, {Kara},
  {Miller-Jones}, {Hung}, {Bright}, {Roth}, {Blagorodnova}, {Huppenkothen},
  {Yan}, {Ofek}, {Sollerman}, {Frederick}, {Ward}, {Graham}, {Fender},
  {Kasliwal}, {Canella}, {Stein}, {Giomi}, {Brinnel}, {van Santen}, {Nordin},
  {Bellm}, {Dekany}, {Fremling}, {Golkhou}, {Kupfer}, {Kulkarni}, {Laher},
  {Mahabal}, {Masci}, {Miller}, {Neill}, {Riddle}, {Rigault}, {Rusholme},
  {Soumagnac}, \& {Tachibana}}]{vanVelzen:2019ApJ}
{van Velzen}, S., {Gezari}, S., {Cenko}, S.~B., {et~al.} 2019, \apj, 872, 198,
  \dodoi{10.3847/1538-4357/aafe0c}

\bibitem[{{van Velzen} {et~al.}(2021){van Velzen}, {Gezari}, {Hammerstein},
  {Roth}, {Frederick}, {Ward}, {Hung}, {Cenko}, {Stein}, {Perley}, {Taggart},
  {Foley}, {Sollerman}, {Blagorodnova}, {Andreoni}, {Bellm}, {Brinnel}, {De},
  {Dekany}, {Feeney}, {Fremling}, {Giomi}, {Golkhou}, {Graham}, {Ho},
  {Kasliwal}, {Kilpatrick}, {Kulkarni}, {Kupfer}, {Laher}, {Mahabal}, {Masci},
  {Miller}, {Nordin}, {Riddle}, {Rusholme}, {van Santen}, {Sharma}, {Shupe}, \&
  {Soumagnac}}]{vanVelzen:2021ApJ}
{van Velzen}, S., {Gezari}, S., {Hammerstein}, E., {et~al.} 2021, \apj, 908, 4,
  \dodoi{10.3847/1538-4357/abc258}

\bibitem[{{Villar} {et~al.}(2018){Villar}, {Nicholl}, \&
  {Berger}}]{Villar:2018ApJ}
{Villar}, V.~A., {Nicholl}, M., \& {Berger}, E. 2018, ApJ, 869, 166,
  \dodoi{10.3847/1538-4357/aaee6a}

\bibitem[{{Zabludoff} {et~al.}(2021){Zabludoff}, {Arcavi}, {La Massa},
  {Perets}, {Trakhtenbrot}, {Zauderer}, {Auchettl}, {Dai}, {French}, {Hung},
  {Kara}, {Lodato}, {Maksym}, {Qin}, {Ramirez-Ruiz}, {Roth}, {Runnoe}, \&
  {Wevers}}]{Zabludoff2021SSRv}
{Zabludoff}, A., {Arcavi}, I., {La Massa}, S., {et~al.} 2021, \ssr, 217, 54,
  \dodoi{10.1007/s11214-021-00829-4}

\end{thebibliography}
\bibliographystyle{aasjournal}



\end{document}